\documentclass[twocolumn]{aastex62}
\pdfoutput=1 
\usepackage{amsmath,amstext}
\usepackage[T1]{fontenc}
\usepackage[figure,figure*]{hypcap}
\usepackage{graphicx}
\usepackage{subfigure}
\usepackage{needspace,enumitem}

\newcommand{\affilcarnegie}{The Observatories of the Carnegie Institution for Science, 813 Santa Barbara St., Pasadena, CA 91101, USA}
\newcommand{\affiltoronto}{Department of Astronomy and Astrophysics, University of Toronto, 50 St. George Street, Toronto, Ontario, M5S 3H4 Canada}

\begin{document}
\shorttitle{Thirty Years of Radio Observations of Type Ia SN 1972E and SN 1895B}
\shortauthors{Cendes et al.}

\title{Thirty Years of Radio Observations of Type Ia SN 1972E and SN 1895B: Constraints on Circumstellar Shells}

\correspondingauthor{Yvette Cendes}
\email{yvette.cendes@cfa.harvard.edu}

\author{Y. Cendes}
\affiliation{Dunlap Institute for Astronomy and Astrophysics University of Toronto, Toronto, ON M5S 3H4, Canada}
\affiliation{\affiltoronto}
\affiliation{Leiden Observatory, PO Box 9513, 2300 RA Leiden, The Netherlands}
\affiliation{Center for Astrophysics \textbar{} Harvard \& Smithsonian, Cambridge, MA 02138, USA}
\author{M. R. Drout}
\altaffiliation{CIFAR Azrieli Global Scholar}
\affiliation{\affiltoronto}
\affiliation{\affilcarnegie}
\author{L. Chomiuk}
\affiliation{Center for Time Domain and Data Intensive Astronomy, Department of Physics and Astronomy, Michigan State University, East Lansing, MI 48824, USA}
\author{S. K. Sarbadhicary}
\affiliation{Center for Time Domain and Data Intensive Astronomy, Department of Physics and Astronomy, Michigan State University, East Lansing, MI 48824, USA}

\begin{abstract}
We have imaged over 35 years of archival Very Large Array (VLA) observations of the nearby (d$_{\rm{L}}$ $=$ 3.15 Mpc) Type Ia supernovae SN\,1972E and SN\,1895B  between 9 and 121 years post-explosion. No radio emission is detected, constraining the 8.5 GHz luminosities of SN\,1972E and SN\,1895B to be L$_{\nu,8.5\rm{GHz}}$ $<$ 6.0 $\times$ 10$^{23}$ erg s$^{-1}$ Hz$^{-1}$ 45 years post-explosion and L$_{\nu,8.5\rm{GHz}}$ $<$ 8.9 $\times$ 10$^{23}$ erg s$^{-1}$ Hz$^{-1}$ 121 years post-explosion, respectively. These limits imply a clean circumstellar medium (CSM), with $n$ $<$ 0.9 cm$^{-3}$ out to radii of a few $\times$ 10$^{18}$ cm, if the SN blastwave is expanding into uniform density material.  Due to the extensive time coverage of our observations, we also constrain the presence of CSM shells surrounding the progenitor of SN\,1972E.  We rule out essentially all medium and thick shells with masses of 0.05$-$0.3 M$_\odot$ at radii between $\sim$10$^{17}$ and 10$^{18}$ cm, and thin shells at specific radii with masses down to $\lesssim$0.01 M$_\odot$. These constraints rule out swaths of parameter space for a range of single and double degenerate progenitor scenarios, including recurrent nova, core-degenerate objects, ultra-prompt explosions and white dwarf (WD) mergers with delays of a few hundred years between the onset of merger and explosion. Allowed progenitors include WD-WD systems with a significant ($>$ 10$^{4}$ years) delay from the last episode of common envelope evolution and single degenerate systems undergoing recurrent nova---provided that the recurrence timescale is relatively short and the system has been in the nova phase for $\gtrsim$10$^{4}$ years, such that a large ($>$ 10$^{18}$ cm) cavity has been evacuated. Future multi-epoch observations of additional intermediate-aged Type Ia SNe will provide a comprehensive view of the large-scale CSM environments around these explosions.
\end{abstract}


\section{Introduction} 
\label{sec:intro}

Type Ia supernovae (SNe) are caused by the explosion of a carbon-oxygen white dwarf \citep[WD;][]{Nomoto1982}.  They have become an important cornerstone of cosmological distance calculations as "standardizable candles" for 
measuring the expansion of the universe
via their measured luminosity distances as a function of redshift \citep{Riess1998,Perlmutter1999}.  However, despite their importance, debates still remain regarding both the progenitor systems and explosion mechanism of Type Ia SNe \citep[e.g.][]{Maoz2014}.

There are two broad scenarios in which a carbon-oxygen WD can explode as Type Ia SNe, and both involve binary systems \citep{Hillebrandt2000, Wang2018}.  The first is the single degenerate (SD) scenario, in which the WD accretes material from a non-degenerate stellar companion \citep{Nomoto1984,Thielemann1986,Holmbo2018}.  The second is the double degenerate (DD) scenario, where the secondary companion is also a WD \citep{Webbink1984,Iben1984,Maoz2014,Liu2018}.
The term "double degenerate" is broad and currently encompasses multiple combinations of progenitor binary systems and explosion mechanisms, including direct collisions \citep{Kushnir2013}, mergers \citep[][]{Shen2012}, and double detonations due to accretion from a helium WD companion \citep{Shen2013,Glasner2018}.  It is also debated whether Type Ia SN can only be produced near the Chandrasekhar Mass (M$_{\rm{Ch}}$), or if sub-M$_{\rm{Ch}}$ WDs can also produce normal Type Ia SNe while undergoing double detonations or violent mergers \citep{Woosley2011,Kromer2010,Shen2018}.  Some observations show evidence for a population of sub-M$_{\rm{Ch}}$ explosions \citep[e.g.][]{Scalzo2019}.

One strategy to shed light on these open questions is to search for circumstellar material (CSM) surrounding Type Ia SNe. The CSM is produced by the pre-explosion evolution of binary system---including winds, outbursts and episodes of mass transfer---and can therefore reflect the nature of the SN progenitor. However, for decades searches for CSM around Type Ia SNe in the X-ray and radio have yielded non-detections \citep[][]{Panagia2006,Hancock2011,Margutti2012,Chomiuk2012,Russell2012,Margutti2014,Chomiuk2015}, implying low-density environments. Most of these observations were taken within a few hundred days of the SN explosion, constraining the density of the CSM at distances $\lesssim$ 10$^{16}$ cm from the progenitor star. Of these, observations of three nearby events---SN\,2011fe, SN\,2014J, and SN\,2012cg---have constrained the pre-explosion mass-loss rates of the progenitor systems to $\dot{M} < 10^{-9} M_{\odot}$ yr$^{-1}$, ruling out all but the lowest mass SD systems \citep{Margutti2012,Chomiuk2012,Margutti2014,Chomiuk2015}. At the same time, larger samples of more distant events systematically rule out winds from more massive or evolved stellar companions \citep[][]{Russell2012,Chomiuk2015}.

In recent years, however, other types of observations have painted a more complex picture of the CSM surrounding Type Ia SNe.  First, a new class of SNe (SNe Ia-CSM) spectroscopically resemble SNe Ia but have strong hydrogen emission lines \citep{Silverman2013}. This has been interpreted as the SN shockwave interacting with a significant amount of CSM ($\sim$few $M_{\odot})$ located directly around the explosion site (distributed out to radii of $\sim$10$^{16}$ cm). SNe Ia-CSM are rare, and the most nearby (SN 2012ca; d$_{\rm{L}}$ $\sim$ 80 Mpc) is the only Type Ia SN detected in X-rays to date \citep{Bochenek2018}. 

Additionally, blue-shifted Na I D absorbing material has been detected in some normal Type Ia SNe spectra, which is interpreted as CSM surrounding the SNe that has been ionized \citep{Patat2007,Blondin2009,Sternberg2011,Maguire2013}.  Modeling has indicated the material is not distributed continuously with radius, but is more likely located in shell-like structures at radii $\geq 10^{17}$ cm \citep{Chugai2008}.  Such absorbing material is estimated to have a total mass of up to $\sim1 M_{\odot}$, and is thought to be present in $\geq$20\% of SNe Ia in spiral galaxies \citep{Sternberg2011}.  Most recently, \citet{Graham2019} reported evidence of CSM interaction surrounding SN 2015cp at $\sim730$ days post-explosion, consistent with a CSM shell that contains hydrogen at distances $\geq$10$^{16}$ cm, and \citet{Kollmeier2019} reported the detection of H$\alpha$ in a late-time nebular spectrum of ASASSN-18tb, interpreted as the signature of CSM interaction.

Despite these intriguing results, constraints on the CSM surrounding Type Ia SNe at radii $\gtrsim$10$^{17}$cm have been relatively sparse. These distances can be probed by radio observations obtained between $\sim$5 and 50 years post-explosion. These timescales have typically been neglected because the deepest constraints on the presence of a stellar wind density profile can be made in the first $\sim$year post-explosion. However, if an uniform density medium is present, deeper limits on CSM would be possible via radio observations at greater times post-SN, as the shockwave continues to interact with the ambient material \citep{Chevalier1998}.  Additionally, if multiple observations are taken over the course of several years, the presence of CSM shells at a range of radii can be probed.

On even longer time scales ($\sim$100 years) radio observations can yield information on the CSM density and structure as a SN transitions to the SN remnant (SNR) stage. In our own galaxy, young Type Ia SNRs have been observed in radio wavelengths. For example, G1.9+0.3, was first discovered by the Very Large Array (VLA) and is estimated to be between 125 and 140 years old \citep{Reynolds2008}.  Additionally, Kepler's SNR is radio bright $\sim$400 years after the explosion \citep{DeLaney2002}.  However, whether this emission is due to interaction with CSM ejected by the progenitor system, or the interstellar medium (ISM), is still debated.  In contrast, \citet{Sarbad2017} made deep radio images of the SN 1885A area in the Andromeda Galaxy (M31; 0.785 $\pm$ 0.025 Mpc distant). The resulting upper limits constrain SN 1885A to be fainter than G1.9+0.3 at a similar timescale of $\sim$120 years post-explosion, placing strict limits on the density of the ambient medium and the transition to the SNR stage.  This appears to favor a sub-M$_{\rm{Ch}}$ model for the explosion.

While observations of SNe within our Local Group (e.g. SN 1885A) can provide the deepest \emph{individual} limits on the CSM density surrounding the progenitors of Type Ia SNe, the number of Type Ia SNe with ages $\lesssim$100 years is limited. Therefore, in order to build up a statistical sample of intermediate-aged SNe, we must look to galaxies farther afield.  In this paper, we have compiled over 30 years of radio observations of NGC\,5253 for this purpose. NGC 5253 offers an ideal example for such studies because (i) it has hosted two Type Ia SNe in the past $\sim$150 years (SN 1972E and SN 1895B), (ii) it is located at very close proximity (d$=$3.15 Mpc; \citealt{Freedman2001}), and (iii) it has been observed with the  historic VLA and upgraded Karl G.\ Jansky VLA multiple times between 1981 and 2016.  
Such a data set over so many years allows us to probe the density of the CSM out to large radii from the SNe, constrain the presence of CSM shells, and provide insight into various progenitor scenarios for Type Ia SNe.  

This paper is structured as follows. In Section \ref{sec:background}, we summarize information known on SN 1895B and SN 1972E.  In Section \ref{sec:observations}, we describe 30 years of archival radio observations of these systems.  In Section \ref{sec:results}, we use these observations to place deep limits on the density of a uniform ambient medium and the presence of CSM shells surrounding SN\,1972E and SN\,1895B at radii between 10$^{17}$ and 10$^{18}$ cm.  In Section \ref{sec:discussion}, we discuss these results in the context of multiple Type Ia SN progenitor scenarios, and the future of SN\,1972E and SN\,1895B as they transition to the SNR stage.

\section{Background: SN 1895B and SN 1972E}
\label{sec:background}

Two independent Type Ia SNe, SN 1895B and SN 1972E, occurred within a century of each other in the nearby blue compact dwarf galaxy, NGC 5253. NGC 5253 is located within the M83/Centaurus A Group, and throughout this work we adopt the Cepheid distance of 3.15 Mpc from \citet{Freedman2001}\footnote{This distance includes a metallicity correction factor.}. NGC 5253 is currently undergoing a starburst phase with a compact, young star forming region at its center \citep{Monreal2010}, thought to be triggered by an earlier interaction with M83 \citep{vandenbergh1980}.

SN 1895B (J2000 Coordinates: RA $=$ 13:39:55.9, Dec $=$ $-$31:38:31) was discovered by Wilhelmina Fleming on December 12, 1895 from a spectrum plate taken on July 18, 1895 \citep{Pickering1895}. Throughout this manuscript, we adopt the discovery date as the explosion epoch for our analysis, although the explosion likely occurred some days earlier. Three direct image plates and one spectrum plate taken within the first five months of the SN are available. Re-analysis of these plates with a scanning microdensiometer have resulted in a light curve that is consistent with a normal Type Ia SN $\sim$15 days after maximum light \citep{Schaefer1995}. From this analysis, it is estimated that SN\,1895B peaked at a visual magnitude of $<8.49\pm0.03$ mag.  

Significantly more information is available for SN 1972E, which was the second-brightest SN of the 20th century.  Discovered on May 13, 1972 (J2000 coordinates: RA $=$ 13:39:52.7, Dec $=$ $-$31:40:09), SN\,1972E was identified just prior to maximum light \citep{Leibundgut1991}, peaked at a visual of 8.5 mag and was observed for 700 days after initial discovery \citep{Kirshner1975,Ardeberg1973,Bolton1974}. As with SN 1895B, we adopt the discovery date as the explosion date for the analysis below\footnote{We note that differences of $\sim$1 month in adopted explosion epoch will not influence our results, as our observations take place tens to 100 years after the explosion.}. 
The exquisite late-time coverage of SN\,1972E at optical wavebands played a key role in our understanding of the link between Type Ia SNe and nucleosynthesis \citep{Trimble1982} as it was shown that the energy deposition during the optical-thin phase was consistent with the radioactive decay of $^{56}$Ni and $^{56}$Co \citep{Axelrod1980}. SN\,1972E is now considered an archetype for Type Ia SN, and was one of the events used to define the spectroscopic features of ``Branch normal'' events \citep{Branch1993}.

\begin{figure}
	\includegraphics[width=\columnwidth]{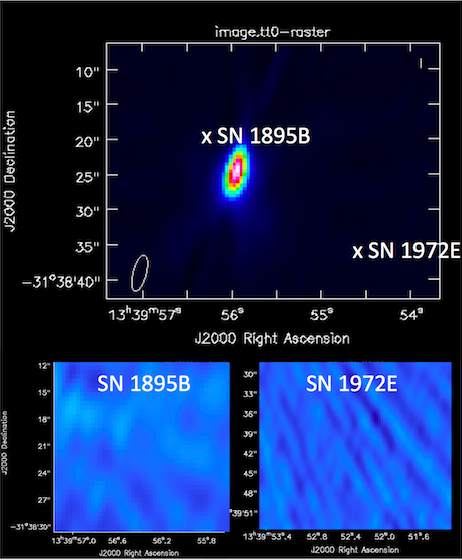}
    \caption{\emph{Top:} Radio image of NGC~5253 from a December 2016 VLA observation at 8.35 GHz, with the positions SN\,1972E and SN\,1895B marked. The bright central radio source in NGC 5253 is a compact star forming region in the galaxy core \citep{Monreal2010}. The synthesized beam is drawn as an ellipse in the lower left corner. \emph{Bottom:} Close-ups of the regions surrounding each SN.}
    \label{fig:radio-image}
\end{figure}

After the initial optical light faded, neither SN 1895B nor SN 1972E has been detected at any wavelength.  Observations of NGC 5253 with the \emph{Chandra} X-ray observatory yielded non-detections at the locations of both SNe \citep{Summers2004}. In the radio, two upper limits for the SNe have been published to date, which are listed in Table \ref{tab:observations}, below. \citet{Cowan1982} observed both SN 1895B and SN 1972E with the VLA for 3 hours at 1.45 GHz in April 1981. They report non-detections with upper limits of 0.9 mJy for both SNe. Subsequently, \citet{Eck2002}, reported upper limits on the radio flux from both SNe of 0.15 mJy based on 9.1 hours of VLA data obtained in November 1984, also at 1.45 GHz. Modeling these limits assuming a CSM with a $\rho \propto r^{-2}$ density profile, \citet{Eck2002} find upper limits on mass-loss rates of the progenitor systems of SN 1972E and SN 1895B of $<8.60\times10^{-6}$ M$_{\odot}$ yr$^{-1}$ and $<7.2\times10^{-5}$ M$_{\odot}$ yr$^{-1}$, respectively. These mass-loss rate estimates, which assumed wind speeds of 10 km s$^{-1}$, are not strongly constraining in the context of Type Ia SN progenitors, ruling out only a few specific Galactic symbiotic systems \citep{Seaquist1990}. 

\begin{deluxetable*}{lccccccr}
\tabletypesize{\small}
\tablecaption{Observation Details for Archival VLA Data \label{tab:observations}}
\tablewidth{0pt}
\tablehead{
\colhead{Observation} & \colhead{Project} & \colhead{Configuration} &
\colhead{Integration} & \colhead{Central Freq.} & \colhead{Receiver} &
\colhead{Bandwidth} & \colhead{Reference\tablenotemark{a}} \\ 
\colhead{Date} & \colhead{Code} &
\colhead{} & \colhead{(hr)} & \colhead{(GHz)} & \colhead{} &
\colhead{(MHz)} & \colhead{}
}
\startdata
Apr.\ 1981\tablenotemark{b} & N/A & BnA & 3.0 & 1.45 & L & 12.5 & (1) \\
Nov.\ 1984\tablenotemark{b}& AB0305 & A & 9.1 & 1.45 & L & 25 & (2) \\
\hline
Oct.\ 13, 1991\tablenotemark{b} & AB0626 & A & 1.36 & 8.40 & X& 50 & This Work \\ 
Feb.\ 18, 1999\tablenotemark{b} & AN0081 & D& 3.60 & 8.30 & X& 25 & This Work \\ 
May 5, 2012 & 12A-184 & CnB & 1.16 & 5.85 & C & 2048 & This Work \\
Mar.\ 23, 2016 & TDEM0022 & C & 0.66 & 9.00 & X & 4096 & This Work \\ 
Dec.\ 16, 2016 & 16B-067 & A & 0.75 & 8.35 & X & 4096 & This Work \\
\enddata
\tablenotetext{a}{(1) \citet{Cowan1982}; (2) \citet{Eck2002}}
\tablenotetext{b}{Historical VLA observations}
\end{deluxetable*}

These two SNe are worthy of further study at radio wavelengths for several reasons. First, at 3.15 Mpc, SN\,1972E and SN\,1895B are among the closest known extragalactic SNe. Second, while radio observations of SNe years after explosion are generally not constraining in the content of a $\rho \propto r^{-2}$ wind environments, even comparatively shallow limits can provide useful constraints on the presence of a constant density CSM \citep{Sarbad2017} and low-density CSM shells \citep{Harris2016}---physical models that were not considered in the analysis of \citet{Eck2002}. Third, NGC 5253 has been observed multiple times by the VLA since 1984, and these observations are currently in the VLA archive. This gives us the unique opportunity of being able to set limits at multiple epochs for two SNe, as the shockwave has traversed a wide range of radii---and potentially, CSM environments.

\section{Observations and Data Reduction}
\label{sec:observations}

\subsection{VLA Observations}

For our study, we examined all archival VLA observations of the galaxy NGC~5253. While over 85 observations of NGC 5253 have been obtained since 1979, the location of SN\,1972E \citep[approximately 56" west and 85" south of the nucleus of NGC 5253][]{Jarrett1973} is too far to be visible in higher frequency images centered on the galaxy core. As a result, we initially restrict ourselves to 24 observations that contain either SN\,1895B or SN\,1972E within their primary beam, and occurred in C and X bands (4-12 GHz). 

Subsequently, we further restrict ourselves to observations that can provide constraints on constant density CSM surrounding the SNe, as described by the model outlined in Section~\ref{sec:model}. In particular, while a higher density CSM will lead to brighter overall radio  emission, it will also cause the SN to enter the Sedov-Taylor phase (and therefore fade at radio wavelengths) at an earlier epoch. Thus, in the context of this physical model, there is a maximum radio luminosity that can be achieved at a given time post-explosion. This translates to a minimum image sensitivity that must be achieved for a given intermediate-aged SN. For the cases of SN\,1972E and SN\,1895B, we find that we require radio images with RMS noise less than 85 mJy/beam. After performing a number tests with historical VLA data of NGC~5253, we find that observations with total on-source integration times less than 20 minutes do  not meet this threshold. After applying these cuts, we are left with two historical (pre-2010) VLA observations in addition to the observations published in \citet{Cowan1982} and \citet{Eck2002}, and three observations taken with the upgraded Karl G.\ Jansky VLA (post-2012).

The information for each observation including date, project code, exposure time, configuration, frequency, and band are shown in Table \ref{tab:observations}. 
Overall, these observations provide constraints on the radio luminosity from SN 1972E and SN 1895B between 9$-$44 years and 86$-$121 years post-SN, respectively.

\subsection{Data Reduction and Imaging}

All VLA data were analyzed with the Common Astronomy Software Applications (CASA; \citealt{McMullin2007}). For the 2012 and 2016 data, taken with the ungraded VLA, CASA tasks were accessed through the python-based \texttt{pwkit} package\footnote{available at: https://github.com/pkgw/pwkit} \citep{Williams2017}, while historical data was reduced manually. We flagged for RFI using the automatic AOFlagger \citep{Offringa2012}. After calibration, we imaged the total intensity component (Stokes I) of the source visibilities, setting the cell size so there would be 4$-$5 pixels across the width of the beam. All data was imaged using the CLEAN algorithm \citep{Cornwell2008}, and for post-2010 data we utilize mfsclean \citep{Rau2011} with nterms $=$ 2. Due to the large distance of SN\,1972E from the galaxy center (and thus image pointing) we also image using the w-projection with wprojplanes $=$ 128. Finally, images were produced setting robust to 0 and for all observations, we used the flux scaling as defined by \citet{Perley2017}.

\begin{deluxetable*}{llccccc}
\tabletypesize{\small}
\tablecaption{Radio Observations of SN\,1972E and SN\,1895B \label{tab:density}}
\tablehead{
\colhead{Supernova} & \colhead{Obs.\ } & \colhead{Time Since} & \colhead{Central} & \colhead{RMS} & \colhead{Luminosity} &\colhead{Density} \\
\colhead{} & \colhead{Date} & \colhead{Explosion\tablenotemark{a}} & \colhead{Freq.} & \colhead{Noise} & \colhead{Upper Limit} & \colhead{Upper Limit\tablenotemark{b}} \\
\colhead{} & \colhead{(UT)} & \colhead{(yrs)} & \colhead{(GHz)} & \colhead{($\mu$Jy/beam)} & \colhead{(ergs/s/Hz)} & \colhead{(cm$^{-3}$)}}
\startdata
1895B     & Apr.\ 1981 & 85.8 & 1.45 & 900 & 3.2E+25 & 4.2 \\
          & Nov.\ 1984 & 89.3 & 1.45 & 150 & 5.3E+24 & 1.0 \\
          & Oct.\ 1991 & 96.3 & 8.40 & 820 & 2.9E+25 & 16 \\
          & Feb.\ 1999 & 103.6 & 8.30 & 33 & 1.2E+24 & 1.1 \\
          & May 2012 & 116.9 & 5.85 & 33 & 1.2E+24 & 0.8 \\
          & Mar.\ 2016 & 120.7 & 9.00 & 77 & 2.7E+24 & 2.1 \\
          & Dec.\ 2016 & 121.5 & 8.35 & 25 & 8.9E+23 & 0.8 \\
\hline	
1972E     & Apr.\ 1981 & 8.9 & 1.45 & 900 & 3.2E+25 & 14 \\
          & Nov.\ 1984 & 12.5 & 1.45 & 150 & 5.3E+24 & 2.6 \\
          & Oct.\ 1991 & 19.4 & 8.40 & 270 & 9.6E+24 & 15 \\
          & Feb.\ 1999 & 27.8 & 8.30 & 26 & 9.2E+23 & 1.7 \\
          & May 2012 & 40.0 & 5.85 & 26 & 9.2E+23 & 1.0 \\
          & Mar.\ 2016 & 43.9 & 9.00 & 40 & 1.4E+24 & 2.0 \\
          & Dec.\ 2016 & 44.6 & 8.35 & 17 & 6.0E+23 & 0.9  
\enddata
\tablenotetext{a}{Assuming the explosion epochs adopted in Section~\ref{sec:background}.}
\tablenotetext{b}{Assuming a constant CSM density, n$_0$, and the fiducial model described in Section~\ref{sec:model}.}
\end{deluxetable*}

For all observations, the center of the radio image is dominated by the bright central radio source in NGC~5253 located at RA $=$ $13\rm{h}39\rm{m}55.96\rm{s}$ and Dec.\ $=$ $-31^{\circ}38^{\prime}24.5^{\prime\prime}$ (J2000; \citealt{Beck1996}).  An example images can be seen in Figure \ref{fig:radio-image}, with the positions of SN\,1972E and SN\,1895 marked for reference.

\subsection{Flux Limits}
\label{sec:flux-limits}

We did not detect radio emission at the location of either SN\,1895B or SN\,1972E in any of our images.  To obtain flux upper limits, we measured the RMS noise at the locations of the SNe using the \texttt{imtool} program within the \texttt{pwkit} package \citep{Williams2017}. These values are listed in Table \ref{tab:density}. Throughout this manuscript, we will assume 3$\sigma$ upper limits radio flux from SN\,1972E and SN\,1895B. 
In general, the upper limits obtained on the flux from SN\,1972E were a factor of $\sim$2$-$3 deeper than for SN\,1895B. This primarily due to that fact that SN\,1895B occurred significantly closer to the radio-bright center of the galaxy (see Figure~\ref{fig:radio-image}). The deepest individual flux limits for both SNe were provided by the December 2016 observation, with 3$\sigma$ upper limits of F$_{\nu}$ $<51\ \mu$Jy/beam and F$_{\nu}$ $<75\ \mu$Jy/beam for SN\,1972E and SN\,1895B, respectively.

\section{Results}
\label{sec:results}

\subsection{Radio Luminosity Limits: Comparison to Previously Observed SNe and SNRs}
\label{sec:SNR-compare}

\begin{figure*}[!ht]
\begin{center}
\includegraphics[width=.9\textwidth]{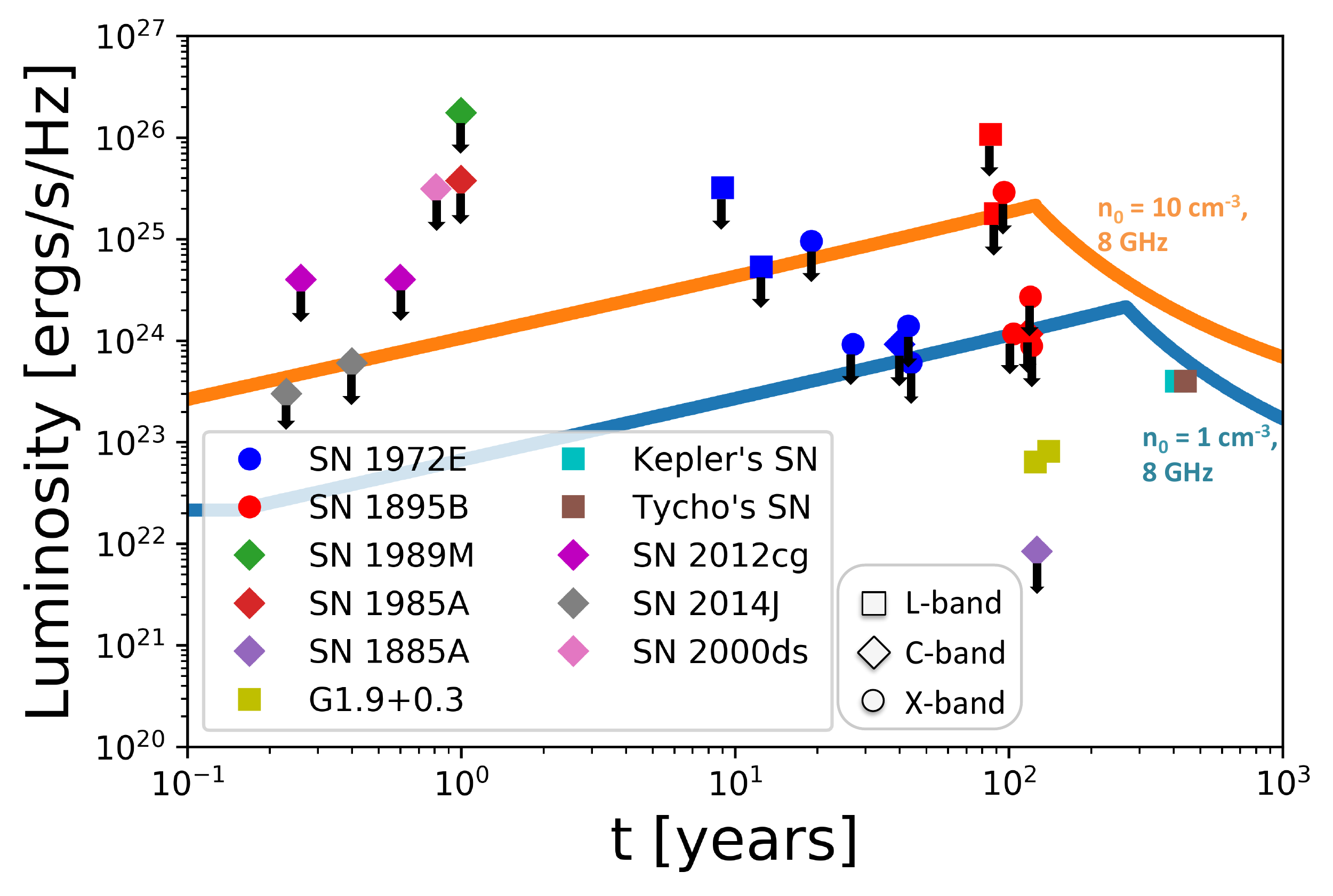}
\end{center}
\caption{Radio luminosity upper limits for the intermediate-aged Type Ia SN\,1972E (blue) and SN\,1895B (red) spanning three decades using data from this work (see Table \ref{tab:observations} and previous observations. \citep{Cowan1982,Eck2002}). Also shown, for comparison, are observed radio luminosities and luminosity upper limits (black arrows) for Galactic SNRs and other extagalactic Type Ia SNe for a range of times post-SN \citep{Chomiuk2015,Sarbad2017,Green2014,Condon1998}. We have distinguished the different observed frequency bands present in this data set as different symbols: squares correspond to L-band (1-2 GHz), diamonds to C-band (4-8 GHz), and circles to X-band (8-12 GHz) observations. For illustrative purposes, we have included solid lines to represent two potential model radio light curves expected for a SN blast wave expanding into a uniform medium with a density of 1 cm$^{-3}$ (blue) and 10 cm$^{-3}$ (orange), assuming our baseline S17 model described in Section~\ref{sec:model}. See Table~\ref{tab:density} for the precise density limits that can be derived from each point. }
\label{fig:upper-limits}
\end{figure*}

Upper limits on the radio luminosity to each SNe, computed using a distance of 3.15 Mpc to NGC~5253, are listed in Table~\ref{tab:density}. We find limits ranging from $\lesssim$3 $\times 10^{25}$ erg s$^{-1}$ Hz$^{-1}$ in 1981 to $\lesssim$6 $\times 10^{23}$ erg s$^{-1}$ Hz$^{-1}$ in 2016. These limits are shown in Figure~\ref{fig:upper-limits}, along with observations of previously observed SNe and SNRs for comparison. Each SN or SNR is plotted in a different color, while symbols indicate the frequency of each observation. Upper limits are designated by black arrows.

Figure \ref{fig:upper-limits} demonstrates the unique timescales and luminosities probed by SN\,1972E and SN\,1895B. In one of the most thorough reviews of radio emission from Type Ia SNe to date \citet{Chomiuk2015} provided observations of 85 Type Ia SNe within 1 year post-explosion. The deepest limits cited in \citet{Chomiuk2015} correspond to luminosities of $\sim$ 3$-$6 $\times$ $10^{23}$ erg/s/Hz for SN 2014J between 84 and 146 days post-explosion, and $\sim$ 4$-$6 $\times$ $10^{24}$ erg/s/Hz for SN 2012cg between 43 and 216 days post-explosion. These are comparable to the limits obtained for SN\,1972E and SN\,1895B, but at a significantly shorter time post-explosion. In Figure~\ref{fig:upper-limits}, we plot the Type Ia SNe with the deepest luminosity limits obtained between 3 months and 1 year post-explosion \citep{Chomiuk2015,Panagia2006}.

While observations of SNe and SNRs within the Milky Way and other Local Group galaxies can provide deeper constraints on the radio luminosity from Type Ia SNe, such observations have typically been obtained at longer timescales post-explosion. This is demonstrated in Figure~\ref{fig:upper-limits}, where we also plot a radio upper limit for SN 1885A in M31 and observed radio luminosities for the Galactic SNRs G1.9+0.3, Tycho, and Kepler, all associated with events of thermonuclear origin \citep{deVauc1985,Fesen2016,Reynolds2008,RuizLapuente2004,Reynolds2007}. 

By co-adding VLA observations in the 4-8 GHz frequency range, \cite{Sarbad2017} produced a deep radio image with RMS noise of 1.3 $\mu$Jy/ beam at the location of SN 1885A in M31. Some radio emission with 2.6$\sigma$ confidence is also present, but the association with SN 1885A for this emission is uncertain due to the large amount of diffuse radio emission in the central regions of M31 where the SN is located. The resulting luminosity upper limit of 8.5$\times 10^{21}$ erg s$^{-1}$ Hz$^{-1}$ at 127 years post-explosion is approximately two orders of magnitude deeper and at timescales just beyond those probed by SN\,1895B. In comparison, the Galactic SNR G1.9+0.3 was detected at 1.4 GHz with a flux of 0.74 $\pm$ 0.04 Jy in 1993 \citep{Condon1998}, and 0.935 $\pm$ 0.047 Jy in 2008 \citep{Green2008}, corresponding to ages of $\sim$125$-$140 years post-explosion \citep{Reynolds2008,Green2008}. Based on a high absorbing column density in observed X-ray observations, \citet{Reynolds2008} place the distance to G1.9+0.3 to be $\sim$8.5 kpc, with corresponding radio luminosities of $\sim10^{23}$ erg/s/Hz. Finally, the Catalog of Galactic Supernova Remnants \citep{Green2014}, lists 1 GHz fluxes of 56 Jy and 19 Jy for Tycho's SNR and Kepler's SNR, respectively. At estimated distances of 2.8 kpc \citep{Kozlova2018} and 6.4 kpc \citep{Reynoso1999}, respectively, these translate to radio luminosities of $\sim 5 \times 10^{23}$ erg/s/Hz. However, we emphasize that these SNe are over 400 years old, and have transitioned to the SNR phase.

Given that the observed luminosities of these Galactic intermediate-aged Type Ia SNe/SNRs are below the luminosity upper limits obtained for SN\,1972E and SN\,1895B, we also calculate the flux densities that they would be observed with at the distance of NGC\,5253. We find that the observed flux densities of G1.9+0.3-like, Kepler-like, and Tycho-like SNRs would be $\sim$2 $\mu$Jy, $\sim$15 $\mu$Jy, and $\sim$ 26 $\mu$Jy in NGC~5253, respectively. These flux levels for Kepler's and Tycho's SNR are within the sensativity limits that can be achieved through dedicated JVLA observations, and the implications for the future evolution of SN\,1972E and SN\,1895B are discussed in Section~\ref{sec:discussion}, below.

\subsection{Constraints on a Uniform Density CSM}

The radio emission from a SN expanding into a relatively low density medium is described by a synchrotron spectrum. As the shockwave expands into the CSM, electrons are accelerated to relativistic speeds and interact with shock-amplified magnetic fields \citep{Chevalier1982,Chevalier2006}. Here, we use a quantitative model for the radio luminosity from a SN blast wave expanding into a constant density CSM and our luminosity upper limits to place constraints on the density of the media surrounding the progenitors SN\,1972E and SN\,1895B.

\subsubsection{Radio Light Curve Model}
\label{sec:model}

We adopt the radio luminosity model outlined in \citet[][S17 hereafter, see their Appendix A]{Sarbad2016}, based on the radio synchrotron formalism of \citet{Chevalier1998}. This model self consistently treats the evolution of the SN from early (ejecta dominated) to late (Sedov-Taylor) phases, and is therefore ideal for the intermediate-aged SNe considered here. While we refer the reader to S17 for a complete model description, we summarize salient features here.

The luminosity of the radio emission from a Type Ia SN will depend on the density profiles of the outer SN ejecta and CSM, the ejecta mass (M$_{\rm{ej}}$) and kinetic energy (E$_{\rm{K}}$) of the SN explosion, the power spectrum of the relativistically accelerated electrons, and the fraction of post-shock energy contained in amplified magnetic fields and relativistic electrons ($\epsilon_{\rm{B}}$ and $\epsilon_e$, respectively). S17 use standard model assumptions in many cases: adopting a power-law density profile with a ``core-envelope'' structure for the SN ejecta as defined by \citet{Truelove1999} with $\rho \propto v^{-n}$
and $n=10$ in the outer ejecta \citep{Matzner1999}, a constant density CSM, and a distribution of relativistic electrons of the form $N(E) \propto E^{-p}$. However, S17 deviate from standard assumptions in their treatment of the magnetic field amplification. 

In most analytic models of SN radio light curves, $\epsilon_{\rm{e}}$ and $\epsilon_{\mathrm{B}}$ are free parameters, assumed to be constant. This is generally considered to be one of the most significant uncertainties in converting observed radio luminosities to CSM densities \citep{Horesh2012,Horesh2013}. In contrast, S17 develop a new parameterization for $\epsilon_{\mathrm{b}}$, as a scaling of the Alfven Mach number of the shock and the cosmic ray acceleration efficiency, based on the results of numerical simulations of particle acceleration \citep{Caprioli2014}. $\epsilon_{\mathrm{B}}$ is therefore determined as a function of time and equipartition is not assumed. As a result, the models of S17 contain five free parameters: $p$, $\epsilon_e$, $M_{\rm{ej}}$, and $E_{\rm{K}}$, and $n_0$ (the density of the CSM).

Given their ages and the analytic models for SN blast wave dynamics of \citet{Truelove1999}, SN\,1972E and SN\,1895B should still be in the free-expansion (ejecta-dominated) phase during the VLA observations described above. During this phase, the radius and velocity of the forward shock can be described by:
\begin{equation}
R_s = (1.29\ \mathrm{pc})\ t_2^{0.7}\ E_{\mathrm{51}}^{0.35}\ n_0^{-0.1}\ M_{\mathrm{ej}}^{-0.25}
\label{eq:radius}
\end{equation}
and
\begin{equation}
v_s = (8797\ \mathrm{km/s})\ t_2^{-0.3}\ E_{\mathrm{51}}^{0.35}\ n_0^{-0.1}\ M_{\mathrm{ej}}^{-0.25}
\label{eq:velocity}
\end{equation}
where $t_2 = t/(100\ \rm{yrs})$, is the time post-explosion, $E_{\rm{51}} = E/(10^{51}\ \rm{ergs})$ is the kinetic energy of explosion, $M_{\rm{ej}} = M/ (1\  \rm{\rm{M_{\odot}}})$ is the ejecta mass, and $n_0$ is the ambient medium density in units of 1 cm$^{-3}$.

Using these relations, and equations A1-A11 in S17, we can then calculate the radio luminosity of a Type Ia SN interacting with a uniform density CSM under the assumption that the resulting synchrotron emission is optically thin and the forward shock will dominate the radio luminosity\footnote{Please note corrections to these equations provided in the erratum \citet{Sarbadhicary2019}.}.
These assumptions hold for the low density ambient media we consider here. 

In Figure~\ref{fig:upper-limits} we plot example S17 light curves for two CSM densities (1 and 10 cm$^{-3}$), assuming a fiducial ``baseline'' model with $M_{\rm ej} =$ 1.4 M$_{\odot}$, $E_{\rm K} = 10^{51}$ erg, $p=3$, and $\epsilon_e$ = 0.1. The latter two values are widely adopted in the literature and are motivated by radio observations of SNe and gamma-ray bursts \citep{Chevalier2006}. However, we emphasize that both $p$ and $\epsilon_e$ may vary based on the source population and $\epsilon_e$, in particular, is subject to significant uncertainty. Observations of young SNRs, such as Tycho, are consistent with a very small $\epsilon_e$ ($\lesssim$ 10$^{-4}$; \citealt{Morlino2012}, \citealt{Berezhko2006}, \citealt{Berezhko2009}), while the luminosity function of older SNRs in local galaxies requires an intermediate value ($\epsilon_e$ $\approx$ 10$^{-3}$; S17). Similarly, while young radio SNe are often consistent with $p=3$ \citep{Chevalier2006}, the spectral index of young SNRs is usually in the range of $p=2.0-2.4$\citep{Dubner2015}. We have chosen our baseline values for $p$ and $\epsilon_e$ both because SN\,1972E and SN\,1895B should still be in the ejecta-dominated phase, and to allow for direct comparison to the observational results of \cite{Chomiuk2015} and the hydrodynamic models of SN-CSM shell interaction described in Section~\ref{sec:shell}. Effects of varying these parameters will be examined below.

From the baseline S17 models presented in Figure~\ref{fig:upper-limits} it is clear the predicted radio luminosity increases steadily during the free-expansion phase---over a timescale of centuries---thus allowing later observations to place deeper constraints on the density of the ambient medium. This is in sharp contrast to a $\rho \propto r^{-2}$ wind environment, where the predicted radio luminosity fades with time as a result of the decreasing density \citep[see, e.g.][]{Chomiuk2015}. In the uniform CSM scenario, the radio light curve peaks a few hundred years after SN, around the Sedov time, and subsequently the radio luminosity declines throughout the Sedov-Taylor phase (S17).

\subsubsection{Limits on Uniform Density CSM}
\label{sec:applied}
\label{sec:density}

\begin{figure}[h]
  \begin{center}
    \subfigure[Expected luminosity for various densities for our ``baseline'' model ($M_{ej} = 1.4\ M_\odot$, E$_{\rm{K}} = 10^{51}$ erg, and $\epsilon_e = 0.1$, with $n_0$ ranging from $0.1-50$ cm$^{-3}$).]{\label{fig:a}\includegraphics[width=\columnwidth]{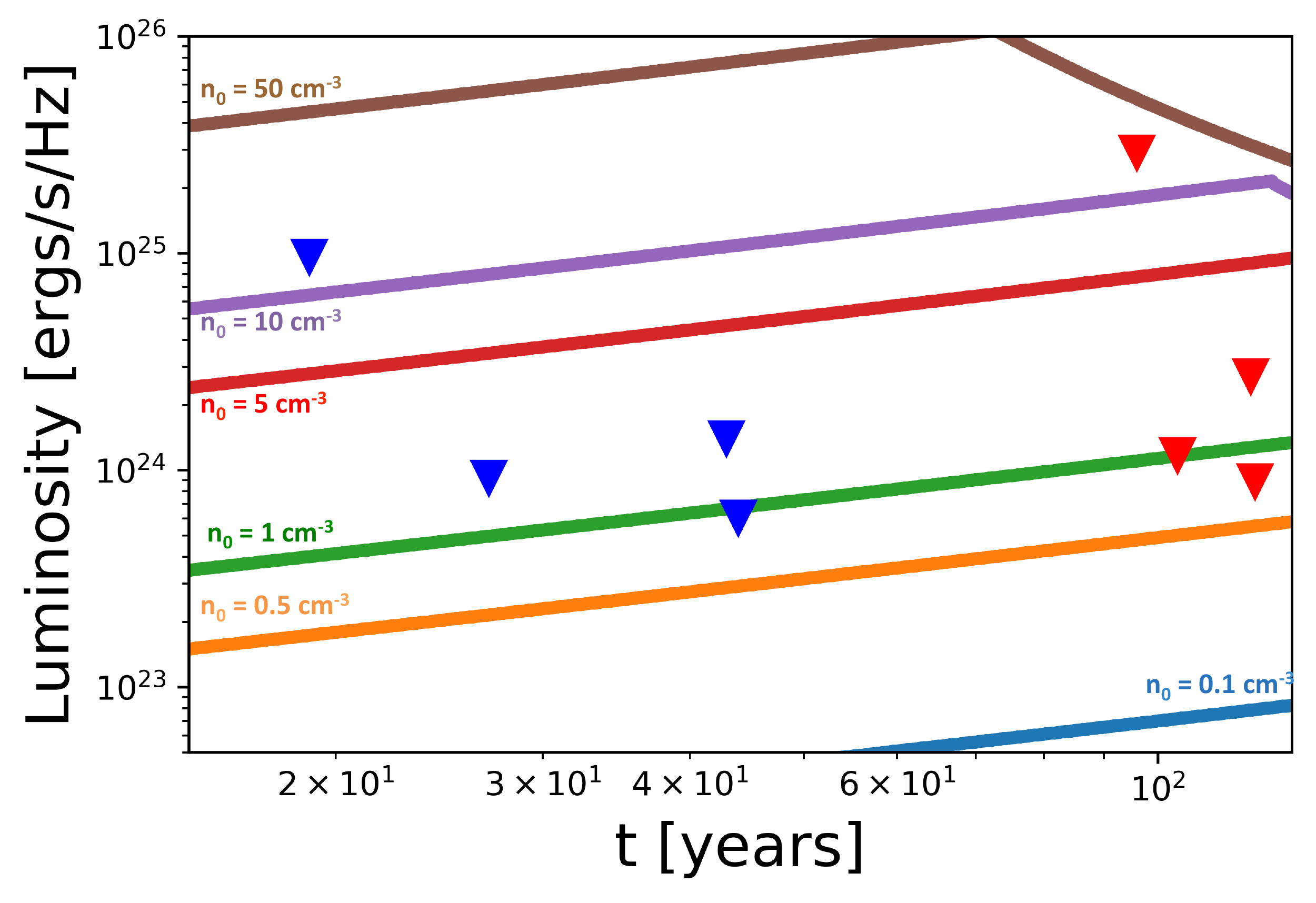}}
    \subfigure[Same as \ref{fig:a}, but $M_{ej} = 0.8\ M_\odot$.]{\label{fig:b}\includegraphics[width=\columnwidth]{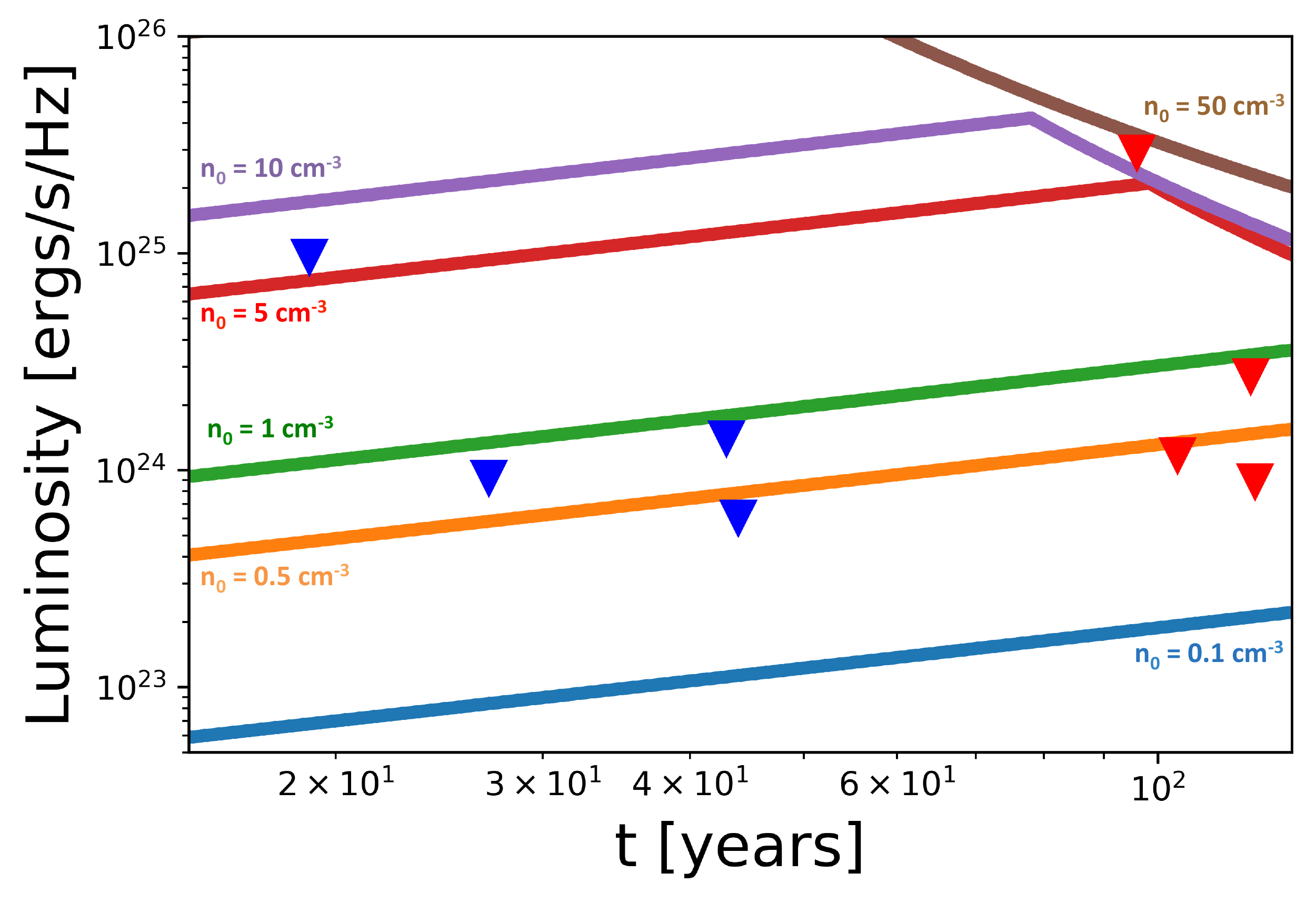}} \\
    \subfigure[Same as \ref{fig:a}, but with $\epsilon_e = 0.01$.]{\label{fig:c}\includegraphics[width=\columnwidth]{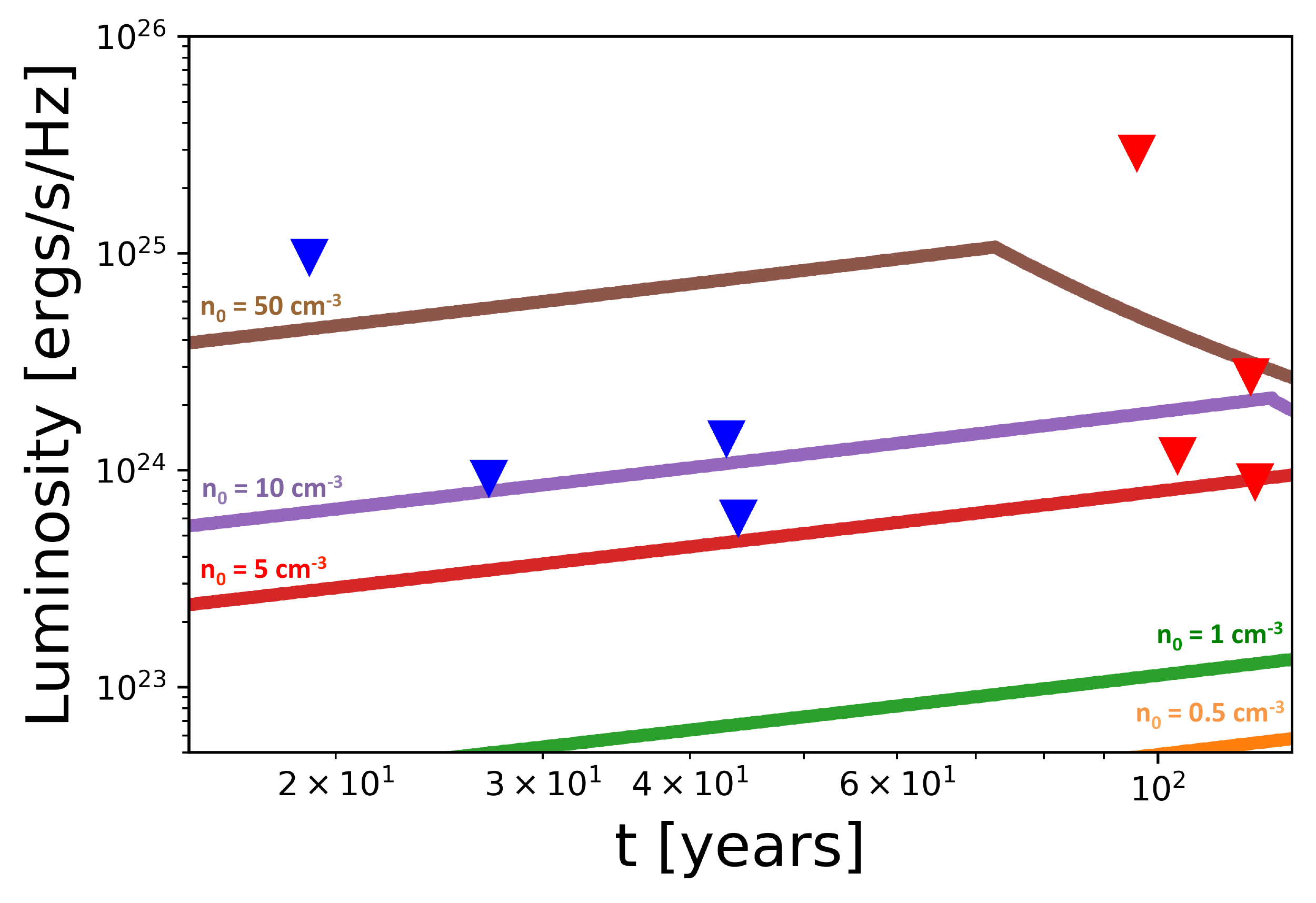}}\\
  \end{center}
  \caption{Expected 8 GHz radio luminosity over time for S17 models if we vary $n_0$, $M_{ej}$, or $\epsilon_{e}$. X-band upper limits for SN\,1972E (blue) and SN\,1895B (red) are provided for comparison. See text for details.}
  \label{fig:varied}
\end{figure}

\begin{figure}[!t]
\includegraphics[width=\columnwidth]{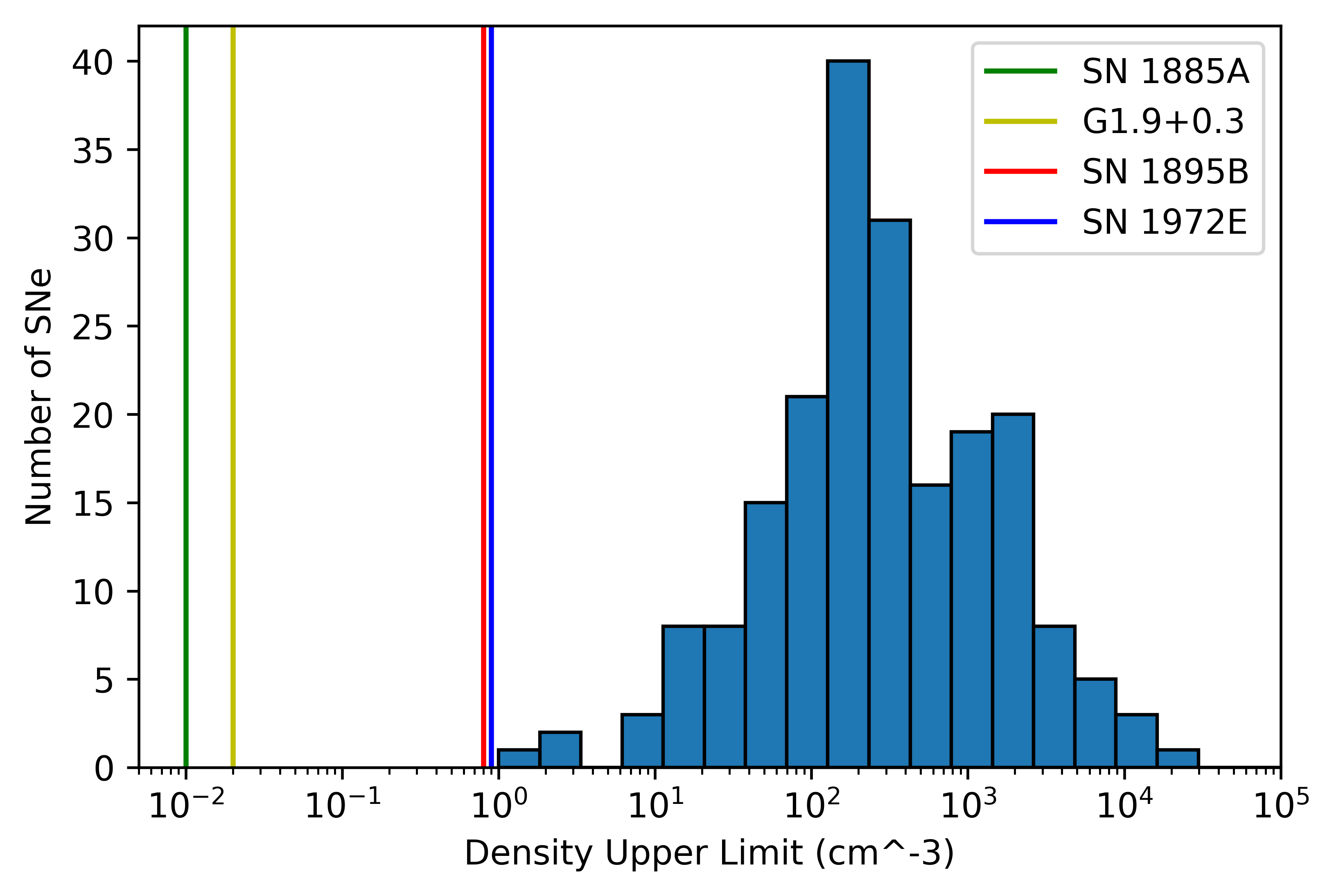}
\caption{A histogram of the uniform density CSM upper limits for $\sim$200 radio observations of 85 Type Ia SNe reported in \citet{Chomiuk2015}, compared to the deepest limits found in this work for SN\,1972E and SN\,1895B (red and blue vertical lines, respectively). Also shown is a density upper limit for SN\,1885A (green) and a density measurement for G1.9+0.8 (yellow) calculated based on the luminosities from \citet{Sarbad2017} and assuming our baseline model described in Section~\ref{sec:model}.}
\label{fig:cumulative}
\end{figure}

We have applied the radio model of S17 to the luminosity upper limits derived for SN\,1972E and SN 1895B (Section~\ref{sec:flux-limits}; Figure~\ref{fig:upper-limits}) in order to place limits on the density of any uniform medium surrounding the SNe. In Table~\ref{tab:density} we list the density upper limits that result when assuming our baseline model described above ($M_{\rm{ej}} =$ 1.4 M$_{\odot}$, $E_{\rm{K}}=$ 10$^{51}$ erg, $p = 3$, and $\epsilon_e$ = 0.1). For each point, we run a large grid of S17 models and the quoted density upper limit corresponds to the curve which goes directly through the 3$\sigma$ luminosity limit plotted in Figure~\ref{fig:upper-limits}). These density upper limits, which were computed assuming a mean molecular weight of 1.4, range from $\sim$1 to $\sim$15 cm$^{-3}$, depending on the epoch, frequency, and sensitivity of the observation.

In the top panel of Figure~\ref{fig:varied}, we plot example 8 GHz light curves for this baseline model at a range of CSM densities, along with the X-band (8-10 GHz) upper limits for SN\,1972E and SN\,1895B. For both SNe, our deepest constraints on the density of the ambient medium come from the Dec.\ 2016 observations, due to a combination of their deeper sensitivity and longer time post-explosion. Assuming our baseline model, these limits correspond to n$_0$ $\lesssim0.8$ cm$^{-3}$ for SN 1895B, and n$_0$ $\lesssim0.9$ cm$^{-3}$ for SN 1972E. In Figure~\ref{fig:cumulative} we plot these density limits in comparison to those for SN 1885A, SNR G1.9+0.3, and the $\sim$200 observations of 85 extra-galactic Type Ia SN from \citet{Chomiuk2015}. For SN 1885A and G1.9+0.3 we have taken the observed luminosities from \cite{Sarbad2017} and computed new density limits assuming our baseline model, as \citet{Sarbad2017} adopted significantly different values of p$=$2.2 and $\epsilon_e = 10^{-4}$. Given both the small distance to NGC 5253 and the fact that the radio luminosity of a SN expanding into a uniform density CSM will continue to increase over time, we are able to place limits on the CSM density surrounding SN\,1972E and SN\,1895B that are several orders of magnitude lower than the bulk of the population presented in \citet{Chomiuk2015}.

In Figure~\ref{fig:varied}, we also examine the influence on our derived density upper limits if we deviate from our baseline model described above. In the middle panel we plot the 8 GHz light curves that result if we consider an ejecta mass of $M_{ej} = 0.8\ M_{\odot}$, representative of sub-M$_{ch}$ explosions \citep[e.g.,][]{Sim2012}. For these parameters, our best ambient density constraints correspond to n$_0$ $< 0.38$ cm$^{-3}$ (SN\,1972E) and n$_0$ $< 0.31$ cm$^{-3}$ (SN\,1895B). Overall, assuming a sub-M$_{ch}$ explosion yields upper limits on the CSM density that are a factor of $\sim$2.5 more constraining (assuming $E_{\rm{K}}=$ is held fixed at 10$^{51}$ erg). Finally, in the lower panel of Figure~\ref{fig:varied} we highlight the influence of varying the adopted value for $\epsilon_e$. Lowering the value of $\epsilon_e$ by a factor of 10 will yield a predicted luminosity for a given density that is a factor of 10 fainter, and a density constraint for a given luminosity upper limit that is a factor of $\sim$7 weaker (for $p = 3$). If we adopt $\epsilon_e$ $=$ 10$^{-4}$ and p $=$ 2.2 as assumed by \citet[][]{Sarbad2017} when modeling SN\,1885A (based on values consistent with young SNRs), our best ambient density constraints become $\sim$17 cm$^{-3}$ (SN\,1972E) and $\sim$16 cm$^{-3}$ (SN\,1895B). In this case, the impact of a lower adopted $p$ value partially cancels the effect of a dramatically lower $\epsilon_e$.

\begin{figure}
\includegraphics[width=\columnwidth]{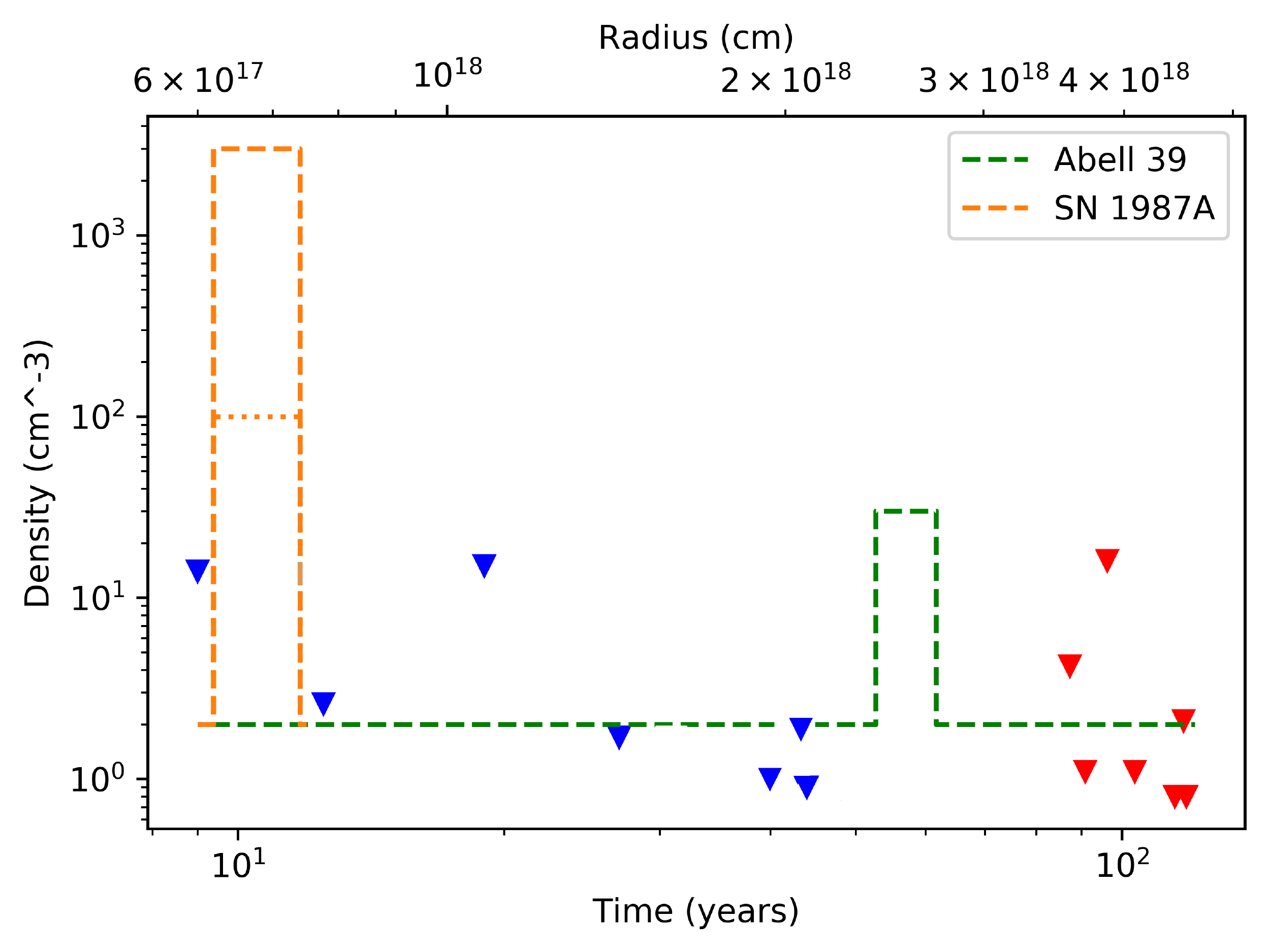}
\caption{The upper limits on density, in cm$^{-3}$, obtained for SN 1972E (blue triangles) and SN 1895B (red) assuming our baseline model.  The top axis shows the radius probed by each observation, assuming a constant density of 1 cm$^{-3}$. For reference, we have also provided a simple density profile for the planetary nebula Abell 39 (green dashed line; \citealt{Jacoby2001}), and the upper and lower limits on the density of the inner ring of SN\,1987A (orange dashed and dotted lines, respectively; \citealt{Mattila2010}).}
\label{fig:dens-abell-39}
\end{figure}

The uniqueness of a data set spanning two decades also allows us to place constraints on the density of the CSM as a function of radius from the progenitor star. In Figure~\ref{fig:dens-abell-39} we plot the uniform density CSM limits obtained for each SN\,1972E and SN\,1895B observation, assuming our baseline S17 model. On the top axis we also provide the radius probed as a function of time, assuming a constant CSM density of 1 cm$^{-3}$. We note that the exact radius probed by each point will vary depending on the density of the CSM (see Equation~\ref{eq:radius}). These densities and radii are similar to those observed in several known CSM shells. For illustrative purposes, we have provided a simple density profile for two such examples: the inner ring of SN 1987A, and the planetary nebula Abell 39. These density profiles should be associated with the top axis of Figure~\ref{fig:dens-abell-39}, which lists the radius from the SN progenitor star.

The radius and density of for inner ring of SN\,1987A are obtained from \citet{Mattila2010}, who provide both upper and lower limits on the ring density (plotted as dashed and dotted orange lines, respectively). For the planetary nebula Abell 39, the radius and density of the shell were obtained via spectroscopic analysis from \citet{Jacoby2001}.  We chose Abell 39 because it is the simplest possible planetary nebula: a one-dimensional projected shell that is used as a benchmark for numerical modeling of these structures \citep{Jacoby2001, Danehkar2012}.  In the case of Abell 39, the shell has a radius of 0.78 pc, a thickness of 0.10 pc, and a density of 30 cm$^{-3}$ \citep{Jacoby2001}.  We have plotted a simple step function where the density is 2 cm$^{-3}$ outside of the shell, consistent with the number density observed within the shell \citep{Toala2016}. This illustrative comparison highlights that even the less sensitive luminosity limits obtained for SN\,1972E and SN\,1895B are useful in constraining the presence of CSM shells. We consider a more detailed model for the radio emission from a SN interacting with CSM shells below.

\subsection{Constraints on the Presence of CSM Shells}\label{sec:shell}

In addition to placing deep limits on the density of uniform CSM, the multi-epoch nature of our radio observations allow us to investigate the possibility of shells of CSM surrounding the progenitors of SN\,1972E and SN\,1895B. Here we outline a parameterized radio light curve model for SN ejecta interacting with spherical shells of finite extent, the applicability of these models to the regimes probed by our observations of SN\,1972E and SN 1895B, and the types of shells that can be ruled out for these systems.

\subsubsection{Radio Light Curve Model: Shell Interaction}
\label{sec:shell-model}

\begin{figure*}[!t]
\begin{center}
\includegraphics[trim={0cm 1.75cm 0cm 12cm},clip,width=\textwidth]{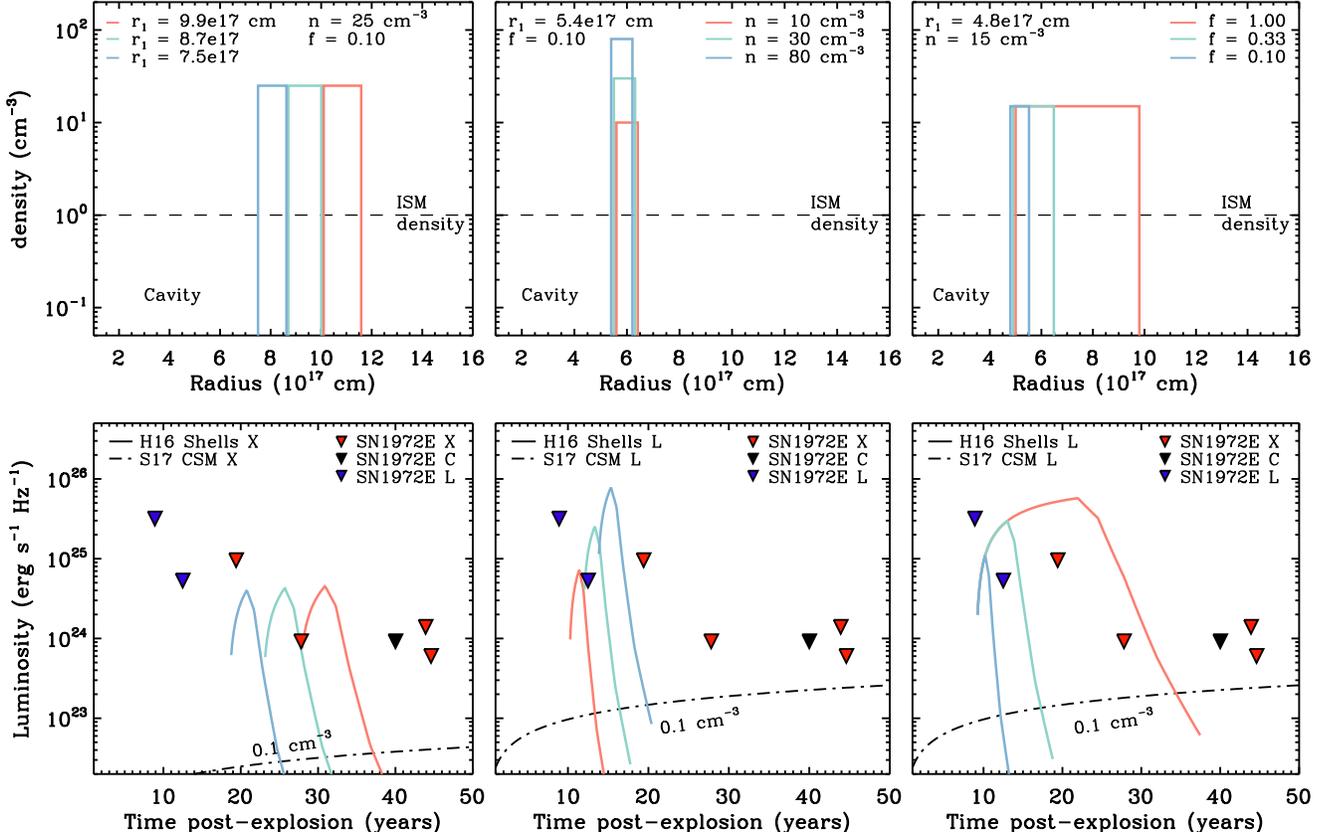}
\end{center}
\caption{Fiducial H16 light curve models for a SN blastwave impacting a constant density CSM shell. The three top panels show representations of the CSM density structure as a function of each of the three free model parameters, varying shell inner radius (r$_1$), shell density (n$_0$), and shell width (f) respectively. A typical ISM density ($\sim$1 cm$^{-3}$; dashed line) is shown as a dashed line for reference. H16 assume a cavity interior to the CSM shell. The three lower panels show the H16 radio light curves that result when a SN blastwave impacts the density structures shown in the panel immediately above them. Radio upper limits from SN 1972E (downward triangles, colors correspond to observed frequency bands) and a 0.1 cm$^{-3}$ constant CSM density radio light curve from S17 (dash-dot line) are shown for comparison. \emph{Left Panels:} Effect of varying shell inner radius. The onset of the resulting radio emission is delayed. \emph{Center panels:} Effect of varying shell density. Higher densities correspond to brighter radio emission and slightly later rise of the radio light curve (see text for details). \emph{Right panels:} Effect of varying the shell fractional width. Radio light curves initially follow the same evolution, but thicker shells yield a longer-lived and brighter radio transient.}
\label{fig:shell-model}
\end{figure*}

To constrain the presence of CSM shells surrounding the progenitors of SN\,1972E and SN\,1895B we use the parameterized light curve models of \cite{Harris2016} (H16, hereafter). H16 model the interaction of expanding SN ejecta with a CSM shell of constant density using the Lagrangian hydrodynamics code of \cite{Roth2015} and compute radio synchrotron light curves based on the gas property outputs of these simulations. While these models can be run for a wide variety of ejecta and CSM configurations, for ease of parameterization, H16 also produced a set of fiducial models for a M$_{\rm{ej}}$ = M$_{\rm{Ch}}$ = 1.38 M$_\odot$ and E$_{\rm{K}}$ = 10$^{51}$ erg Type Ia SN, with a physical set-up that is based off of the self-similar formalism of \citet{Chevalier1982}.

Specifically, for this fiducial model set, H16 adopt power law density profiles for both the SN ejecta and CSM, and set the initial conditions of the simulations such that the initial contact discontinuity radius equals the contact discontinuity radius at the time of impact from \citet{Chevalier1982}. Following \citet{Chevalier1994} and \citet{Kasen2010}, the SN ejecta is defined by a broken power law with shallow and steep density profiles ($\rho$ $\propto$ r$^{-1}$ vs.\ $\rho$ $\propto$ r$^{-10}$) for material interior and beyond a transition velocity, $v_{t}$, respectively. The CSM is defined as a shell with a finite fixed width, $\Delta$R, and constant density, n.

In constructing radio synchrotron light curves from the outputs of this fiducial set of models H16 assume that $\epsilon_e = \epsilon_B = 0.1$ and that that the accelerated electrons possess a power-law structure with respect to their Lorentz factor, $\gamma$, of the form $n(\gamma) \propto \gamma^{-p}$ with $p=3$. It is also assumed that the radio emission is dominated by the forward shock and that the resulting emission is optically thin to synchrotron self-absorption, assumptions that were shown to be valid for their model set.

With these assumptions, H16 find a ``family'' of resulting radio synchrotron light curves that can be defined by three key parameters:
\begin{itemize}
    \item[$r_{1}$:] the inner radius of the CSM shell.
    \item[$n$:] the density of the CSM shell.
    \item[$f$:] the fractional width of the CSM shell ($\Delta$R/$r_{1}$).
\end{itemize}
H16 provide analytic expressions describing radio light curves as a function of these three parameters.

In Figure~\ref{fig:shell-model} we plot the resulting radio light curves (lower panels) for various CSM shells (top panels) as each of $r_1$, $n$, and $f$ are varied individually. Also shown, for context, are the luminosity upper limits measured for SN\,1972E and the radio light curve for a 0.1 cm$^{-3}$ constant density CSM from S17. Overall, the resulting radio light curves are strongly peaked in time, with a rapid decline occurring once the forward shock reaches the outer radius of the CSM shell. For a constant shell density and fractional width (left  panels), adjusting the inner radius of the shell will primarily influence the time of impact and therefore the onset of radio emission. Adjusting the density of the CSM shell (center panels) will primarily influence the peak luminosity of the resulting radio emission---although the onset of radio emission will also be delayed slightly for higher density shells (see below). Finally, increasing the fractional width of the shell (right panels) will increase both the overall timescale and peak luminosity of the resulting radio signature as the interaction continues for a longer time period. Thus, a given observed data point will constrain the presence of a thick shell over a larger range of $r_1$, compared to thin shells with similar densities.

\subsubsection{Applicability to SN 1972E and SN 1895B}

H16 first developed and applied their fiducial models to investigate the case of low-density CSM shells located at radii $\lesssim$ a few $\times 10^{16}$ cm, whose presence would manifest in radio light curves within the first $\sim$1 year post-explosion. We now examine whether the assumptions made in H16 are applicable for CSM shells that would manifest at the timescales of the observations of SN\,1972E and SN\,1895B described above.

\begin{figure}[!t]
\begin{center}
\includegraphics[trim={1cm 0.5cm 2cm 10cm},clip,width=0.5\textwidth]{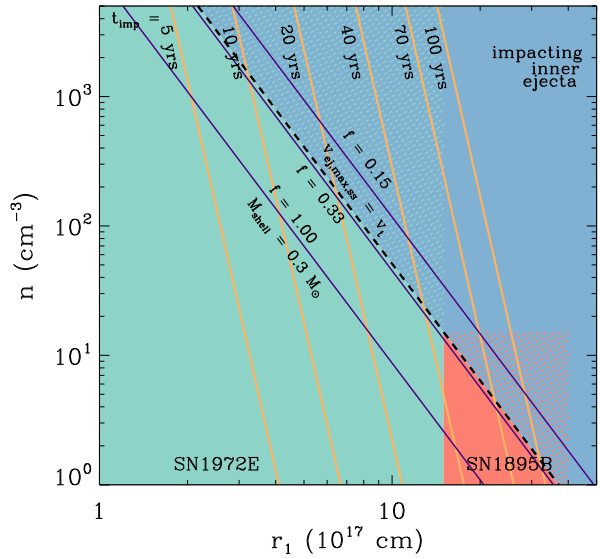}
\end{center}
\caption{Visual representation of the CSM shell inner radii (r$_1$) and densities (n) probed by observations of SN\,1972E and SN\,1895B (aqua and red boxes, respectively) and regions where the model assumptions of H16 are valid. The shaded blue region highlights the parameter space where the assumption that the CSM impacts the outer portion of the SN ejecta is violated. Violet lines indicate densities for which the total shell mass equals the total mass in the outer ejecta ($\sim$0.3M$_\odot$) for fiducial thin, medium, and thick shells. Yellow lines designate the time when the SN ejecta would impact the CSM shell in the models of H16. See text for details.}
\label{fig:H16Phase}
\end{figure}

\begin{figure*}[!t]
\begin{center}
\includegraphics[trim={0cm 1cm 0cm 18cm},clip,width=\textwidth]{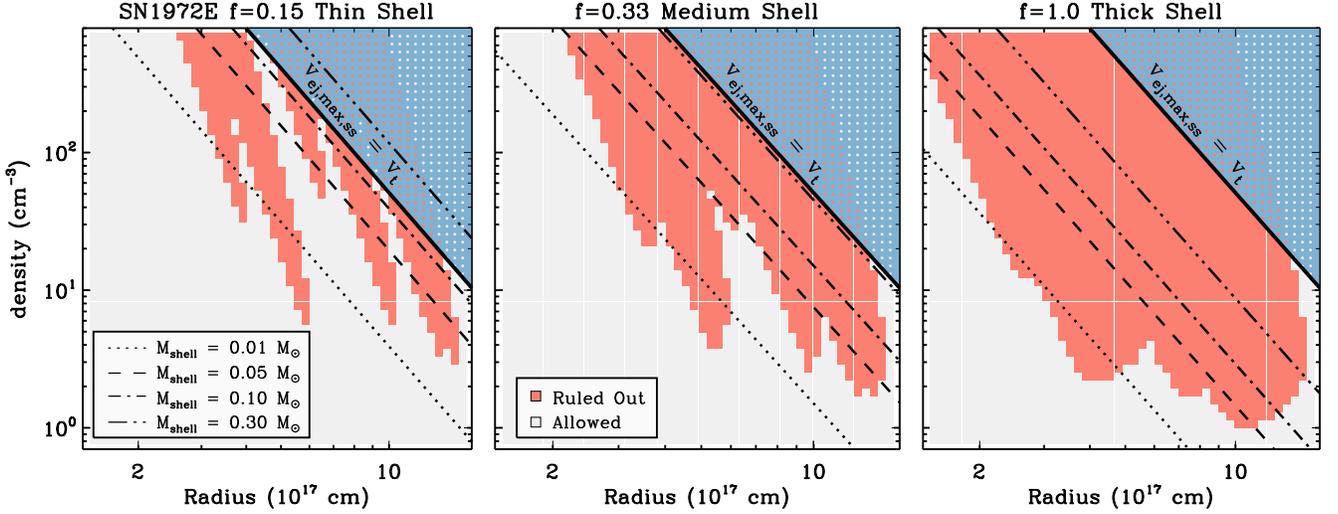}
\end{center}
\caption{Grid of H16 CSM shell models tested against observations of SN\,1972E. Red squares designate the shell radii and densities ruled out for representative thin (left panel), medium (center panel), and thick (right panel) shells. The blue shaded area designates the region where H16 model assumptions are violated. In each panel dotted, dashed, dot-dashed, and triple-dot-dashed lines designate shells with total masses 0.01, 0.05, 0.1, and 0.3 M$_\odot$, respectively. For all shell thicknesses, we can rule out shells with masses down to 0.005$-$0.01 M$_\odot$ for specific radii, and for medium and thick shells our observation exclude the presence of essentially any shell with masses $>$0.05 M$_\odot$ at radii between 10$^{17}$ and 10$^{18}$ cm. See text for further details.}
\label{fig:72ERuledOut}
\end{figure*}

The main assumption that may be violated for the case of shells at the radii probed by the observations of SN 1972E and SN 1895B is that the CSM impacts the \emph{outer} portion of the SN ejecta, which has a steep density profile. For this to hold true, first the total mass swept up by the SN shock prior to impacting the shell should not approach the mass in the outer SN ejecta. For the broken power-law ejecta profile adopted in H16, $\sim$2/9 of the SN ejecta mass is located in the outer ejecta, corresponding to $\sim$0.3 M$_\odot$ for a Chandrasekhar mass explosion. H16 assume that the shell occurs essentially in a vacuum. If we instead assume a low density medium interior to the shell of $<$0.1 cm$^{-3}$ \citep[e.g.][]{Badenes2007} we find that that mass of the internal material swept up should be $\lesssim$0.002 M$_{\odot}$ for the shell radii probed by the observations of SN\,1972E.

Thus, the CSM shell density and radius are the primary determinants of whether the interaction is with the outer SN ejecta. In setting the initial conditions of their simulations, H16 assume that the ``impact'', and hence the beginning of the radio light curve, occurs when the ratio of the CSM and SN ejecta density at the contact discontinuity reaches a specific value ($\rho_{\rm{CSM}} = 0.33\,  \rho_{\rm{ej}}$). This requirement is the cause of the shift in radio emission onset time when considering shells of various densities at a fixed radius. For denser shells, the H16 impact will occur when a slightly denser---more slowly moving---portion of the SN ejecta reaches $r_1$. Thus, at every radius, there is a density that corresponds to 0.33$\,\rho_{\rm{ej,vt}}$ where $\rho_{\rm{ej,vt}}$ is the density of the ejecta at the transition velocity, $v_t$, between the outer and inner density profiles. This is the maximum density of a CSM shell at this radius that does not violate the model assumption that the impact occurs in the outer portion of the SN ejecta. Because the density of the expanding SN ejecta decreases with time, as we consider shells at larger and larger radii, this model assumption will break down for lower and lower densities.

Assuming CSM shells with fractional widths between 0.1 and 1.0, we find that the observations of SN\,1972E and SN\,1895B will probe CSM shells with inner radii ranging between [1$-$15] $\times$ 10$^{17}$ cm and [1.5$-$4.0] $\times$ 10$^{18}$ cm, respectively. In Figure~\ref{fig:H16Phase} we show these ranges in comparison to the model assumption constraints described above. For SN\,1972E, we find that there are large swaths of parameter space that can be probed using the parameterized light curves of H16. However, for SN\,1895B, we find that only shells with very low densities ($\lesssim$ 10 cm$^{-3}$) will not violate model assumptions.

Finally, we note one other requirement based on the assumption that the interaction primarily occurs in the outer SN ejecta: the total mass in the CSM shell should not exceed the total mass in the outer SN ejecta ($\sim$0.3 M$_{\odot}$). Parameter space where this requirement is met and violated are discussed in Section~\ref{sec:shellexcl}, below.

\subsubsection{CSM Shell Models Excluded}
\label{sec:shellexcl}
Finding that the H16 model assumptions are valid over a portion of the parameter space of CSM shells probed by SN\,1972E and SN\,1895B, we run large grids of parameterized light curve models for comparison with our observations. For SN\,1972E, we run 3,200 models for shell radii spanning $r_1 = [1-15] \times 10^{17}$ cm and shell densities spanning $n = 1-16,000$  cm$^{-3}$ ($\sim 2.3 \times 10^{-24}$ to $3.7 \times 10^{-20}$ g cm$^{-3}$). This grid is chosen to encompass the full range of densities that can be probed without violating the the model assumptions described above. For the highest densities considered these models assumptions are only valid at the smallest radii (see Figures~\ref{fig:H16Phase} and ~\ref{fig:72ERuledOut}).
For SN\,1895B we consider 450 models spanning shell radii of $r_1 = [1.6-4] \times 10^{18}$ cm and shell densities of $n = 1-15$ cm$^{-3}$.
For each event, we run models for three representative shell widths, chosen to span the range of astrophysical shells predicted surrounding some putative Type Ia SN progenitors (see Section~\ref{sec:discussion}). Specifically, we consider $f$ values of: 

\begin{itemize}[leftmargin=1.5cm]
    \item[$f=0.15$:] A thin shell based on the based on the observed width of the Abell 39 planetary nebula, and in line with widths predicted for some material swept up in nova outbursts \citep[e.g.][]{Moore2012}.
    \item[$f=0.33$:] A medium thickness shell based on models of ``nova super shells'' \citep{Darnley2019}.
    \item[$f=1.00$:] A representative thick shell.
\end{itemize}

For each combination of $f$, $r_1$, and $n$, we compute the resulting radio light curve at the frequencies of all of our observations and determine whether any of the flux upper limits described above rule out a shell with those parameters. Results from this process for SN\,1972E are shown in Figure~\ref{fig:72ERuledOut}. Shells excluded by the data are displayed in red. For reference, we also plot lines that indicate constant shell masses of 0.01, 0.05, 0.1, and 0.3 M$_\odot$ for each shell thickness. Shells with total masses $>$0.3 M$_\odot$ violate the H16 requirement that the total shell mass be less than the mass in the outer SN ejecta. Regions where the condition that the initial interaction occurs in the outer SN ejecta is violated are also shown in blue.  For the medium-thickness shells considered here, theses two conditions are violated at very similar shell densities, while for thick-shells the constraint that the total shell mass be less than 0.3 M$_\odot$ is the more restrictive requirement (see Figure~\ref{fig:72ERuledOut}; right panel).

Each observed luminosity limit leads to a diagonal line of excluded CSM shells in the density-inner radius plane. This is due to the interplay between the shell density and the onset time of strong interaction that leads to radio emission (see above). For thin shells (left panel) these individual ``tracks'' of excluded models are visible, while for thicker shells (center and right panels) they broaden and overlap. Thus, excluding a complete set of thin shell models for an individual SN progenitor would require higher cadence radio observations than those available for SN\,1972E. In contrast, for thicker shells, we are primarily limited by the depth of individual observations. 

Overall, for SN\,1972E, we can rule out CSM shells down to masses of $\sim$0.01 M$_\odot$ at a range of radii, which vary depending on the shell thickness. We can also rule out the presence of \emph{all} thick shells with masses $\gtrsim$0.05 M$_\odot$ at radii between 1$\times$10$^{17}$ and 1$\times$10$^{18}$ cm, and most medium-width shells of similar mass at radii between 2$\times$10$^{17}$ and 1.5$\times$10$^{18}$ cm. In terms of raw CSM shell density, our deepest limits come between 1 and 1.5 $\times$10$^{18}$ cm, where we can rule out shells with densities between 1 and 3 cm$^{-3}$. 

We emphasize that these radii are larger than those probed by most other observations searching for CSM surrounding Type Ia SNe to date, including time-varying absorption features \citep[e.g.,][]{Patat2007} and late-time optical photometry/spectroscopy \citep[e.g.][]{Graham2019}, which tend to constrain the presence  of CSM around $\sim$10$^{16}$ cm. \citet{Simon2009} do find a radius of $\sim$ 3 $\times 10^{17}$ cm for the material responsible for time-varying Na absorption lines around the Type Ia SN\,2007le. However, the density inferred is much higher ($\sim$10$^{7}$ cm$^{-3}$) and fractional width much narrower ($f \approx 3\times 10^{-4}$) than those considered here, possibly suggesting a clumpy or aspherical CSM. Our observations constrain a unique parameter space of CSM shells.

For SN\,1895B, we find that essentially all of the shell models that would be excluded by the depth and timing of our observations fall in the regime where the H16 assumption that the CSM impacts the outer SN ejecta is violated. However, a few specific exceptions to this exist. For example, we can rule out the presence of an $f=0.33$ medium width shell with a density of 6 cm$^{-3}$ at a radius of $\sim 2 \times 10^{18}$ cm (total shell mass $\sim$0.3 M$_\odot$).  These borderline cases demonstrate that the observations of SN 1895B are likely useful to constrain the presence of shells at these radii, but updated models that include interaction with the dense inner SN ejecta are required for a quantitative assessment.

\section{Discussion}
\label{sec:discussion}

The CSM environment surrounding a Type Ia SN is dependent on pre-explosion evolutionary history of the progenitor system. In this section, we will consider different types of CSM that are both allowed and ruled out by our results (Section \ref{sec:results}), and what they indicate in the context of various Type Ia SN progenitor scenarios.  In Section \ref{sec:constant}, we consider the presence of constant density material, the only material expected in DD scenarios with significant delay times.  We next consider the presence of shells (Section \ref{sec:shells}), as may be expected for SD progenitors if they contain nova shells or planetary nebula and DD progenitors in the case of a prompt explosion post-common envelope.  We also consider the presence of other types of CSM (Section \ref{sec:other}).  Finally, in Section \ref{sec:future}, we make predictions for the future of both SN\,1895B and SN\,1972E as the SNe evolve and future observations are taken.

\subsection{Presence of Constant Density CSM or ISM}
\label{sec:constant}

Our deepest luminosity limits constrain the density of a uniform ambient medium surrounding SN\,1972E and SN\,1895B to be $\lesssim$0.9 cm$^{-3}$ out to radii of $\sim$ 10$^{17}$ $-$ 10$^{18}$ cm. This implies a clean circumstellar environment out to distances 1$-$2 orders of magnitude further than those previously probed by prompt radio and X-ray observations \citep{Chomiuk2012,Margutti2014}. Densities of this level are consistent with the warm phase of the ISM in some galaxies \citep[e.g.][]{Ferriere2001}, and we examine whether our density constraints for SN\,1972E and SN\,1895B are consistent with expectations for the ISM in their local environments within the intensely star-forming galaxy NGC\,5253. 

Using the HI observations of \citet{Kobulnicky1995}, \citet{Summers2004} estimate the ISM density at the location of SN\,1972E, which is $>$1.5 kpc from the central star-forming region, to be $\lesssim$ 1 cm$^{-3}$---comparable to our radio limits. 
In contrast, SN\,1895B exploded $\sim$100 pc from the nucleus of NGC~5253, in a complex region with multiple large stellar clusters (Section \ref{sec:background}). Excluding the dense stellar clusters themselves, \citet{Monreal2010} use IFU spectroscopy with VLT-FLAMES to conclude that the ISM density in this central region is $<$ 100 cm$^{-3}$, and could potentially be 1$-$2 orders of magnitude lower and the explosion site of SN\,1895B, depending on the local distribution of material. Thus, despite some uncertainty, we find that our deepest radio limits constrain the density surrounding SN\,1972E and SN\,1895B to be at levels comparable to, or below, the local ISM at distances of $\sim$ 10$^{17}$ $-$ 10$^{18}$cm. 

Low density media surrounding Type Ia SNe can be achieved through multiple progenitor scenarios. Clean, ISM-like, environments are most commonly evoked for DD models produced by the merger of two WDs. The components of such systems have low intrinsic mass loss rates and current population synthesis models predict that $>$90\% of WD mergers should occur $>$10$^{5}$ years after the last phase of common envelope evolution \citep{Ruiter2013}. Thus, the material ejected during this phase should fully disperse into the ISM at radii beyond 10$^{18}$ cm by the time of explosion. While WD mergers may also pollute the CSM via a number of other physical mechanisms including tidal tail ejections \citep{Raskin2013}, outflows during a phase of rapid mass transfer pre-merger \citep{Guillochon2010,Dan2011} and accretion disk winds in systems that fail to detonate promptly \citep{Ji2013}, this material will be located at radii $<$ a few $\times$ 10$^{17}$ cm, unless there is a significant ($\gtrsim$100 years) delay between the onset of merger and the subsequent Type Ia explosion. In this case, the small amount of material ejected via these mechanisms ($\sim$10$^{-3}$ $-$ 10$^{-2}$ M$_\odot$) will have either dispersed to densities below our measurements or swept up material into a thin shell \citep{Raskin2013}, whose presence will be assessed below. Thus, we conclude that our low inferred densities surrounding SN\,1972E and SN\,1895B are be consistent with expectations for a majority of DD explosions due to WD mergers. 

However, low density ambient media can also be produced by SD and DD Type Ia SN models in which either fast winds or shells of material are ejected from the progenitor system prior to explosion. This high velocity material will subsequently ``sweep-up'' the surrounding ISM, yielding low density cavities surrounding the stellar system \citep[e.g.][]{Badenes2007}. For example, recent hydrodynamical simulations of recurrent nova systems find cavity densities of 10$^{-1}$ $-$ 10$^{-3}$ cm$^{-3}$, far below the density of the ambient ISM \citep{Dimitriadis2014,Darnley2019}. Our radio observations would require a cavity that extends to a few $\times$10$^{18}$ cm. These distances are consistent with the large (r $>$ 10$^{19}$ cm) cavities predicted to be carved by fast accretion wind outflows from the WD surface in some SD models \citep{Hachisu1996}, although such cavities may be inconsistent with observed SNR dynamics \citep{Badenes2007}. In the context of recurrent nova systems such large cavities would require a system that had been undergoing outbursts for $\gtrsim$10,000 years \citep{Dimitriadis2014,Darnley2019}. In the section below, we discuss constraints on the presence of CSM shells surrounding SN\,1972E and SN\,1895B, and thus further implications for this class of progenitor model if a cavity is the source of the clean CSM environments observed.

\subsection{Presence of Shells}
\label{sec:shells}

Several putative progenitor systems for Type Ia SNe predict the presence of shells surrounding the system at distances in the range of those probed by our observations ($\sim10^{17}-10^{18}$ cm). These include both SD and DD systems, with examples of shell creation mechanisms ranging from a recurrent nova to common envelope ejections. In Section~\ref{sec:shellexcl}, we utilized the models of \citet{Harris2016} to explore the basic parameter space of shells that can be constrained and ruled out by our data. Here, we discuss the implications of these results for various progenitor scenarios.

\subsubsection{Recurrent Nova Progenitors}

A recurrent nova is a high mass accreting WD system that undergoes repeating thermonuclear outbursts due to unstable hydrogen burning on its surface, ejecting mass from the system every $\sim 1-100$ yr. The identification of time variable absorption and blue shifted Na I D lines in some Type Ia SNe \citep[][]{Patat2007,Blondin2009,Sternberg2011,Maguire2013} have raised the question of a connection between recurrent novae and Type Ia SNe, particularly in light of the discovery of blue-shifted Na I D lines in the recurrent nova RS Ophiuchi (RS Oph) during outburst \citep{Patat2011,Booth2016}. 

Individual nova eruptions eject a small mass of material (M$_{\rm{ej}}$ $\sim$ 10$^{-7}$ $-$ 10$^{-5}$ M$_\odot$) at high velocities ($v_{\rm{ej}} \gtrsim 3000$ km/s; \citealt{Moore2012,Darnley2019}). However, this material will rapidly decelerate to velocities on the order of tens of km s$^{-1}$ as it sweeps up material from the ISM, CSM, or collides previously ejected shells. The result is a complex CSM structure consisting of of low-density ($n$ $\sim$ 10$^{-1}$ $-$ 10$^{-3}$ cm$^{-3}$) cavities enclosed by a dense outer shell \citep[e.g.][]{Munari1999,Badenes2007}. For a for a $10^4$ year recurring nova phase, such as that seen in RS Oph-like stars, the outer cavity wall predicted to be at radii of $\gtrsim 3\times10^{17}$ cm \citep[e.g.][]{Booth2016,Dimitriadis2014}, within the regime probed by our observations.

The constraints that our observations can provide on the presence of nova shells surrounding SN\,1972E depend primarily on their predicted densities, radii, and thicknesses, which in turn depend on the density of the ambient ISM, the total time the system has been in an active nova phase, and the recurrence timescale between eruptions. Two recent hydrodynamic models for the CSM structure surrounding such systems are presented by \citet{Dimitriadis2014} and \citet{Darnley2019}. The former models nova eruptions with 25, 100, 200 year recurrence timescales expanding into a CSM shaped by winds from a red giant donor star with $\dot{M}$ $=$ 10$^{-6}$ M$_\odot$ and v$_{\rm{w}}$ $=$ 10 km s $^{-1}$. The the latter simulated eruptions with both a shorter recurrence timescale (350 days) and a lower density CSM (shaped by a red giant star with $\dot{M}$ $=$ 2.6 $\times$ 10$^{-8}$ M$_\odot$ and v$_{\rm{w}}$ $=$ 20 km s$^{-1}$). This model was specifically designed to reproduce the CSM surrounding the M31 nova system M31N 2008-12a. M31N 2008-12a is particularly interesting system as it is the most frequently recurring nova known, the WD is predicted to surpass the Chandrasekhar limit in $<$20,000 years \citep{Darnley2017}, and it is surrounded by an observed cavity-shell system with a total projected size of $\sim$ $134 \times 90$ pc \citep{Darnley2019}.

\citet{Dimitriadis2014} find that the density of individual nova ejections expanding into the main cavity depends on the nova recurrence timescale. For longer recurrence times the densities will be higher, as the donor star has additional time to pollute the CSM. For the donor mass-loss rate and recurrence timescales considered by \citet{Dimitriadis2014} these shells are predicted to have densities $\gtrsim$ 10$^{2}$ cm$^{-3}$, while the low density and short recurrence timescale of \citet{Darnley2019} yield individual shell densities below the detection threshold of our observations ($n$ $\lesssim$ 0.1 cm$^{-3}$). However, while our observations can rule out high-density shells from some individual nova eruptions, they are predicted to be too thin ($f$ $\sim$ 0.01; \citealt{Dimitriadis2014}) for our sparse observations to conclusively rule our a system of shells predicted for any specific recurrence time.

In contrast, the outer cavity wall is expected to be thicker. \citet{Darnley2019} find that this ``nova super-remnant shell'' converges a width of $f$ $=$ 0.22 and density approximately 4 times that of the ISM in their simulations ($\sim$ 4 cm$^{-3}$). Our observations can rule out the presence of even these low-density medium-thickness shells at radii between $\sim$ 5 $\times$10$^{17}$ cm and 2 $\times$ 10$^{18}$ cm.  \citet{Darnley2019} find that the outer cavity would be located at these radii for nova systems that have been active for between $\sim$10$^{3}$ and $10^4$ years (having undergone $\sim$1000 $-$ 10,000 total eruptions). For higher density CSM and longer recurrence times, \citet{Dimitriadis2014} find that the cavities will expand more slowly, and thus our observations will rule out older systems.

\subsubsection{Core Degenerate Scenario}

In the core degenerate scenario for Type Ia SNe, a WD companion merges with the hot core of an asymptotic giant branch (AGB) star at the end of a common envelope (CE) or planetary nebula (PN) phase \citep{Kashi2011,Soker2011}. The result of this merger is a massive (M $\gtrsim$ M$_{\rm{Ch}}$), rapidly-rotating, and highly magnetized WD \citep{Tout2008,Kashi2011}, which can subsequently explode as a Type Ia SN. In this scenario, the delay time between the merger and the SN---and hence the location of the CE or PN shell---is primarily set by the spin-down timescale of the merger remnant \citep{Ilkov2012}.   

While originally proposed as a mechanism for prompt explosion after CE ejection (in order to explain Type Ia SN with strong hydrogen emission; \citealt{Livio2003}), a wide range of spin-down timescales are permitted \citep{Lindblom1999, Yoon2005, Ilkov2012}. Based on a number of observational probes, \citet{Tsebrenko2015} have suggested that $\sim$20\% of all Type Ia SN should occur within a PN that ejected within the $\sim$10$^5$ years prior to explosion due to the core-degenerate scenario. Assuming average expansion velocities of tens of km s$^{-1}$, our observations of SN1972E constrain the presence of PN ejected between a few $\times$ 10$^{3}$ and a few $\times$ 10$^{4}$ years prior to explosion. We find we can rule out the presence of roughly Abell 39-like PN (with $n$ $\sim$ 30 cm$^{-3}$ and $f$ $=$ 0.15 at $r_1$ $\sim$ 10$^{18}$ cm) for most of this range of delay times. More broadly, observed PN have masses in the range of $\sim$0.1 M$_\odot$ to 1 M$_\odot$. Our observations rule out most shells with masses between 0.05 M$_\odot$ and 0.3 M$_\odot$ and thicknesses greater than f$=$0.15. Our observations likely also constrain higher mass PN---relevant as the core-degenerate scenario may require massive AGB stars \citep{Livio2003,Kashi2011}---but updated theoretical models, which include the effects of the inner SN ejecta impacting the CSM, are required for quantitative assessment.

\subsubsection{Shell Ejections in DD Progenitors}

There are multiple mechanisms by which DD Type Ia progenitors may also eject shells of material pre-explosion. First, all putative DD progenitor scenarios must undergo at least one episode of CE evolution, in order to yield the requisite tight double WD system  \citep[e.g.][]{Ivanova2013}. For WD merger models, the delay between CE ejection and SN is primarily set by the binary separation post-CE and the gravitational-wave timescale. While current binary population synthesis models predict a majority of WD mergers will occur with a significant delay post-CE, \citet{Ruiter2013} highlight a channel wherein $\sim$ 3.5\% of WD binaries with a massive ($>$0.9 M$_\odot$) primary will merge between 10$^3$ and 10$^{4}$ post-CE. As described above, assuming expansion velocities of a few tens to 100 km s$^{-1}$, our observations of SN\,1972E constrain shells ejected on these timescales. While the CE mass ejection process is uncertain, the total envelop ejected for putative Type Ia progenitors ranges from a few tenths to $\sim$1 M$_\odot$ \citep[e.g.][]{MacLeod2017}. We can rule out most CE shells with masses between 0.05 M$_\odot$ and 0.3 M$_\odot$, unless they are very thin ($f \lesssim 0.1$). Thus, it is unlikely that SN\,1972E underwent and ultra-prompt explosion, although we caution additional theoretical models are required to quantitatively rule out CE shells with masses of $\sim$1 M$_\odot$.

For DD models that are triggered by the detonation of thin surface layer of helium accreted from a low-mass WD companion (the ``double detonation'' model; e.g., \citealt{Woosley1994, Livne1995, Shen2013}), the explosion is predicted to occur between 10$^{8}$ and 10$^{9}$ years after CE \citep{Ruiter2011,Shen2013}. As such, any CE shell will have long since dispersed into the ISM. However, \citep{Shen2013} outline a model whereby such systems can also eject small amounts of hydrogen-rich material (a few $\times$10$^{-5}$ M$_\odot$) at high velocities ($\sim$15000 km s$^{-1}$) in the hundreds to thousands of years before the SN. Analogous to classical novae, this material will sweep up the ISM, forming a cavity and outer shell structure whose properties (mass, radius, thickness) depend on both the evolutionary history of the WD and the ambient ISM density. For ISM densities of 1 cm$^{-3}$, \citep{Shen2013} predict shells with $n$ $\sim$ 5 cm$^{-3}$ and widths of $f$ $\sim$ 0.25 at radii ranging from $r_1$ $\sim$ 5 $\times$10$^{17}$ cm (for older WD progenitors) to $r_1$ $\sim$ 1 $\times$ 10$^{18}$ cm (for younger WD progenitors). Our deepest limits just rule out the presence of such shells around SN1972E, although some intermediate ages are permitted. For sparser ambient ISM densities, such shells would be below our detection limits.

\subsubsection{Tidal Tail Ejections}

In WD-WD merger scenarios, a small amount of material (a few $\times$10$^{-3}$ M$_\odot$) can be ejected in the form of tidal tails, which are stripped from the system just prior to coalescence \citep{Raskin2013}. The ultimate location of this material depends on the delay between the initiation of the merger and the ultimate explosion, and the non-detection of Type Ia SN in prompt (t $\lesssim$ year) radio and X-ray observations have been used to argue for either very short ($\lesssim$ 100 s) or long ($>$ 100 years) delays \citep{Margutti2014,Raskin2013}. For a delay time of $\sim$100 year, \citet{Raskin2013} predict that the tidal tails should appears as a wide ($f = 1$) shell-like structure with a density of $n$ $\sim$ 100 cm$^{-3}$ at a radius of $r_1$ $\sim$ 2 $\times$ 10$^{17}$ cm. Our observations rule out such a CSM structure for SN\,1972E. From this time onward, the tidal material will sweep-up ISM material, decelerating and narrowing in the process. Thus, our observation likely rule out delay times of a few hundred years for this scenario, with the exact range depending on the ISM density and deceleration timescale. \citet{Raskin2013} predict that by 3000 years post-ejection, the tidal material will be located at a radius of $\sim$ 8 $\times$10$^{18}$ cm, well beyond those probed by our observations.

\subsection{Other CSM Structures}
\label{sec:other}

There are several putative Type Ia SN explosion models that predict the presence of CSM, which is neither constant in density nor strictly in the form of shells. Here, we discuss two such cases.

\subsubsection{Stellar Winds}

If the CSM surrounding surrounding the Type Ia SN has a stellar wind-like density distribution ($\rho$ $\propto$ $r^{-2}$), observations from the first $\sim$year post-explosion would provide the deepest constraints on the mass-loss rate of the progenitor system. This density distribution is what is typically expected in SD models that undergo quasi-steady mass-loss due to either winds from a giant (symbiotic) donor star \citep{Seaquist1990}, optically-thick winds from the WD itself during phases of high-accretion \citep{Hachisu1996}, or non-conservative mass-loss through the second Lagrange point during Roche Lobe overflow for some binary configurations \citep{Deufel1999}. In all such cases, emission from the CSM interaction would be strongest in the first days after the SN event when the density of the CSM is highest \citep{Chomiuk2015}. As described in Section~\ref{sec:background}, the deepest limits on the mass-loss rates for SN 1972E and SN 1895B come from the 1984 observations, 12.5 and 8.3 years post-explosion. The constraints of $<8.60\times10^{-6}$ M$_{\odot}$ yr$^{-1}$ and $<7.2\times10^{-5}$ M$_{\odot}$ yr$^{-1}$ (for wind velocities of 10 km s$^{-1}$) rule out a number of Galactic symbiotic systems \citep{Seaquist1990}, but are otherwise unconstraining. We note that these limits depend linearly on the assumed wind speed, and hence for v$_{\rm{w}}$ $>$ 10 km s$^{-1}$ the mass-loss constraints would be even weaker.

\subsubsection{Mass Loss from a Radially Extended Envelope}

\citet{Shen2012} present updated model for the long-term evolution of the remnants of WD mergers, in which the lower mass WD is disrupted and forms a hot radially extended (r $\sim$ 10$^{13}$ cm) envelope around the central remnant rather than an accretion disk. While the final fate of such remnants are debated, it should persist for $\gtrsim$10$^{4}$ years as a carbon burning shell, ignited off-axis, propagates inward to the core.  While they neglect mass loss in their calculations, \citet{Shen2012} note that with a typical escape velocities of 60 km s$^{-1}$, material lost during this phase in the remnant's evolution could reach radii of $\sim$2$\times$10$^{18}$cm, within the radius range probed by our observations.

Subsequently, \citet{Schwab2016} perform updated models and examine the consequences of different mass loss prescriptions on the evolution of such merger remnants. In particular, they note the similarities between the observed properties of these remnants and AGB stars, raising the possibility that a dusty wind may form during an $\sim$5000 year phase in their evolution. Within this context, we note that our observations rule out mass loss on the level observed in extreme AGB stars ($\dot{M}$ $\sim$ 10$^{-4}$ M$_\odot$) out to radii of a few times 10$^{17}$ cm for wind speeds between 10 and 100 km s$^{-1}$. However, \citet{Schwab2016} also find that the temperature of the merger remnant will eventually increase, in a process analogous to PN formation in AGB stars. As a consequence of this evolution, any phase of intense dusty mass loss should cease and the increased UV radiation from the central star could yield an ionized nebulae with total mass of $\sim$0.1 M$_\odot$ at a distance of $\gtrsim$3 $\times$ 10$^{17}$ cm.  Our observations rule out the presence of such shells over a wide range of radii, unless they are very thin ($f$ $<$ 0.1).

\subsection{The Future: SN to SNR Transition}
\label{sec:future}

Our upper limits on the radio luminosity from SN\,1972E and SN\,1895B are consistent with both SN blastwaves expanding into low density CSM environments out to radii of a few $\times$ 10$^{18}$ cm. Assuming a constant density CSM, the radio emission from both events is predicted to continue rising over time (see Figure~\ref{fig:upper-limits}), and we can use our baseline S17 model described in Section~\ref{sec:model} to project their future evolution and thus prospects for subsequent radio detections.
If we assume that no CSM shells are present, and that the SNe are expanding into ambient densities of 0.7 cm$^{-3}$ (just below our Dec.\ 2016 limits; Table~\ref{tab:density}), then both SN would peak at a flux level of $\sim$200 $\mu$Jy (at 5 GHz) $\sim$300 years post-explosion. In this scenario SN\,1972E and SN\,1895B would reach maximum observed brightness in 2272 and 2195, respectively. If we assume that \emph{targeted} VLA observations of each SN could achieve C-band images with RMS noise levels of $\sim$5$-$10 $\mu$Jy (consistent with the sensitivity limits obtained by \citealt{Chomiuk2015}), then both SN1972E and SN\,1895B would currently be detectable at a greater than 5$\sigma$ level.

However, the ambient density surrounding both SNe may be significantly lower than the upper limits found in Section~\ref{sec:results}. In this case the radio light curve would peak at later times and fainter flux levels (S17; See Figure~\ref{fig:upper-limits}). For example, the youngest SN in our own Galaxy, G1.9+0.3, is detected at radio wavelengths at a level consistent ambient density of 0.02 cm$^{-3}$ (\citealt{Sarbad2017}; adjusted for consistency with our baseline S17 model; see Figure~\ref{fig:cumulative}). 
If SN\,1972E and SN\,1895B are expanding into similar CSM environments, then we project that they would peak at 5 GHz flux levels of $\sim$6 $\mu$Jy $\sim$990 years post-explosion. In such a scenario, their current 5 GHz fluxes would be only $\sim$ 1 $\mu$Jy and $\sim$ 2 $\mu$Jy, respectively, and they would never rise above the optimal VLA sensitivity limits described above. This indicates that observations of intermediate-aged Type Ia SNe in nearby galaxies may still be sensitivity limited without significant time investment \citep[10-12 hours; see ][]{Sarbad2017} from current instruments. Limits with future radio telescopes such as the Square Kilometer Array (SKA) and the Next Generation VLA will allow us to study radio emission from Type Ia SNe throughout the local volume, even when they are expanding into low density ($\sim0.1$ cm$^{-3}$) surroundings.

Once additional detections of intermediate aged-SN and young SNRs are made, interpretation of the results will require careful consideration of whether the emission is due to CSM shaped by the progenitor system, or simply the ambient ISM. For example, recent analysis of radio observations of SN\,1885A by \citet{Sarbad2017} conclude the density surrounding the system must be approximately a factor 5 lower density that that surrounding G1.9+0.3. However, they argue that the higher density found for G1.9+0.3 may be due to a higher density in the Milky Way's center---as  compared to M31's---and does not require CSM from the progenitor. Currently, we cannot distinguish between these scenarios based on our data for  SN\,1895B and SN\,1972E.

\section{Summary and Conclusions}

We have conducted a study of the circumstellar environments of the nearby Type Ia SN\,1972E and SN\,1895B by analyzing seven epoch of archival VLA observations from obtained between 1981 and 2016. We do not detect emission from the location of either SN in our data set. The most stringent upper limits on the radio luminosity from each event are L$_{\nu,8.5\rm{GHz}}$ $<$ 8.9 $\times$ 10$^{23}$ erg s$^{-1}$ Hz$^{-1}$ 121 years post-explosion for SN\,1895B and L$_{\nu,8.5\rm{GHz}}$ $<$ 6.0 $\times$ 10$^{23}$ erg s$^{-1}$ Hz$^{-1}$ 45 years post-explosion for SN\,1972E. These imply low-density environments with $n$ $<$ 0.9 cm$^{-3}$ out to radii of a few $\times$ 10$^{18}$ cm --- nearly two orders of magnitude further from the progenitor star than those previously probed by prompt (t $\lesssim$ 1 year) radio and X-ray observations \citep[e.g.][]{Panagia2006,Margutti2012,Chomiuk2012,Russell2012,Margutti2014,Chomiuk2015}. These ambient densities are consistent with progenitor scenarios that produce either ISM-like environments or low-density evacuated cavities out to large distances. 

Given the multi-epoch nature of our dataset, we also investigate the possibility of shells surrounding the progenitor of SN\,19722E. Using the models of H16, we rule out the presence of essentially all medium and thick CSM shells with total masses of 0.05 M$_\odot$ to 0.3 M$_\odot$ located at radii between a few $\times$ $10^{17}$ and a few $\times$ $10^{18}$ cm. We also exclude specific CSM shells down to masses of $\lesssim$ 0.01 M$_\odot$ at a range of radii, which vary depending on the shell thickness (see Figure~\ref{fig:72ERuledOut}). Quantitative assessment of the presence of more massive CSM shells will require updated theoretical models that include the effect of the inner SN ejecta impacting the CSM shell. 

These shell constraints rule out swaths of parameter space for various SD and DD Type Ia SN progenitor models including recurrent nova, core-degenerate objects, ultra-prompt explosions post-CE, shells ejections from CO$+$He WD systems, and WD mergers with delays of a few hundred years between the onset of merger and explosion. Allowed progenitor systems include DD in which the delay from the last episode of CE is long ($>$ 10$^{4}$ yrs) as well as SD models that exhibit nova eruptions---provided the system has a relatively short recurrence timescale and has been in the nova phase for either a short ($\lesssim$100 yrs) or long ($\gtrsim$10$^{4}$ yrs) time. 

It is clear that multi-epoch radio observations of nearby intermediate-aged Type Ia SNe explore useful regions of parameter space for distinguishing between the plethora of theoretical progenitor models. In the future, a statistical sample of such events will provide even more robust discriminating power, as different models predict a range of delay times and hence a variety of locations for CSM material.

\section{Acknowledgements}

We would like to thank C. Harris for their generous assistance in our modeling work and answering our questions and B. Shappee for conversations that inspired this work.  
Support for this work was provided to MRD through Hubble Fellowship grant NSG-HF2-51373, awarded by the Space Telescope Science Institute, which is operated by the Association of Universities for Research in Astronomy, Inc.\ for NASA, under contract NAS5-26555. 
MRD acknowledges support from the Dunlap Institute at the University of Toronto and the Canadian Institute for Advanced Research (CIFAR). 
LC and SS are grateful for the support of NSF AST-1412980, NSF AST-1412549, and NSF AST-1751874.

The National Radio Astronomy Observatory is a facility of the National Science Foundation operated under cooperative agreement by Associated Universities, Inc.  The Dunlap Institute is funded through an endowment established by the David Dunlap family and the University of Toronto.

\software{CASA \citep{McMullin2007}, pwkit \citep{Williams2017}, Astropy \citep{Astropy2013},
Matplotlib \citep{Hunter2007}}


\begin{thebibliography}{}
\expandafter\ifx\csname natexlab\endcsname\relax\def\natexlab#1{#1}\fi
\providecommand{\url}[1]{\href{#1}{#1}}
\providecommand{\dodoi}[1]{doi:~\href{http://doi.org/#1}{\nolinkurl{#1}}}
\providecommand{\doeprint}[1]{\href{http://ascl.net/#1}{\nolinkurl{http://ascl.net/#1}}}
\providecommand{\doarXiv}[1]{\href{https://arxiv.org/abs/#1}{\nolinkurl{https://arxiv.org/abs/#1}}}

\bibitem[{{Ardeberg} \& {de Groot}(1973)}]{Ardeberg1973}
{Ardeberg}, A., \& {de Groot}, M. 1973, \aap, 28, 295

\bibitem[{{Astropy Collaboration} {et~al.}(2013){Astropy Collaboration},
  {Robitaille}, {Tollerud}, {Greenfield}, {Droettboom}, {Bray}, {Aldcroft},
  {Davis}, {Ginsburg}, {Price-Whelan}, {Kerzendorf}, {Conley}, {Crighton},
  {Barbary}, {Muna}, {Ferguson}, {Grollier}, {Parikh}, {Nair}, {Unther},
  {Deil}, {Woillez}, {Conseil}, {Kramer}, {Turner}, {Singer}, {Fox}, {Weaver},
  {Zabalza}, {Edwards}, {Azalee Bostroem}, {Burke}, {Casey}, {Crawford},
  {Dencheva}, {Ely}, {Jenness}, {Labrie}, {Lim}, {Pierfederici}, {Pontzen},
  {Ptak}, {Refsdal}, {Servillat}, \& {Streicher}}]{Astropy2013}
{Astropy Collaboration}, {Robitaille}, T.~P., {Tollerud}, E.~J., {et~al.} 2013,
  \aap, 558, A33, \dodoi{10.1051/0004-6361/201322068}

\bibitem[{{Axelrod}(1980)}]{Axelrod1980}
{Axelrod}, T.~S. 1980, PhD thesis, California Univ., Santa Cruz.

\bibitem[{{Badenes} {et~al.}(2007){Badenes}, {Hughes}, {Bravo}, \&
  {Langer}}]{Badenes2007}
{Badenes}, C., {Hughes}, J.~P., {Bravo}, E., \& {Langer}, N. 2007, \apj, 662,
  472, \dodoi{10.1086/518022}

\bibitem[{{Beck} {et~al.}(1996){Beck}, {Turner}, {Ho}, {Lacy}, \&
  {Kelly}}]{Beck1996}
{Beck}, S.~C., {Turner}, J.~L., {Ho}, P.~T.~P., {Lacy}, J.~H., \& {Kelly},
  D.~M. 1996, \apj, 457, 610, \dodoi{10.1086/176757}

\bibitem[{{Berezhko} {et~al.}(2009){Berezhko}, {Ksenofontov}, \&
  {V{\"o}lk}}]{Berezhko2009}
{Berezhko}, E.~G., {Ksenofontov}, L.~T., \& {V{\"o}lk}, H.~J. 2009, \aap, 505,
  169, \dodoi{10.1051/0004-6361/200911948}

\bibitem[{{Berezhko} \& {V{\"o}lk}(2006)}]{Berezhko2006}
{Berezhko}, E.~G., \& {V{\"o}lk}, H.~J. 2006, \aap, 451, 981,
  \dodoi{10.1051/0004-6361:20054595}

\bibitem[{{Blondin} {et~al.}(2009){Blondin}, {Prieto}, {Patat}, {Challis},
  {Hicken}, {Kirshner}, {Matheson}, \& {Modjaz}}]{Blondin2009}
{Blondin}, S., {Prieto}, J.~L., {Patat}, F., {et~al.} 2009, \apj, 693, 207,
  \dodoi{10.1088/0004-637X/693/1/207}

\bibitem[{{Bochenek} {et~al.}(2018){Bochenek}, {Dwarkadas}, {Silverman}, {Fox},
  {Chevalier}, {Smith}, \& {Filippenko}}]{Bochenek2018}
{Bochenek}, C.~D., {Dwarkadas}, V.~V., {Silverman}, J.~M., {et~al.} 2018,
  \mnras, 473, 336, \dodoi{10.1093/mnras/stx2029}

\bibitem[{{Bolton} {et~al.}(1974){Bolton}, {Garrison}, {Salmon}, \&
  {Geffken}}]{Bolton1974}
{Bolton}, C.~T., {Garrison}, R.~F., {Salmon}, D., \& {Geffken}, N. 1974, \pasp,
  86, 439, \dodoi{10.1086/129627}

\bibitem[{{Booth} {et~al.}(2016){Booth}, {Mohamed}, \&
  {Podsiadlowski}}]{Booth2016}
{Booth}, R.~A., {Mohamed}, S., \& {Podsiadlowski}, P. 2016, \mnras, 457, 822,
  \dodoi{10.1093/mnras/stw001}

\bibitem[{{Branch} {et~al.}(1993){Branch}, {Fisher}, \& {Nugent}}]{Branch1993}
{Branch}, D., {Fisher}, A., \& {Nugent}, P. 1993, \aj, 106, 2383,
  \dodoi{10.1086/116810}

\bibitem[{{Caprioli} \& {Spitkovsky}(2014)}]{Caprioli2014}
{Caprioli}, D., \& {Spitkovsky}, A. 2014, \apj, 794, 46,
  \dodoi{10.1088/0004-637X/794/1/46}

\bibitem[{{Chevalier}(1982)}]{Chevalier1982}
{Chevalier}, R.~A. 1982, \apj, 259, 302, \dodoi{10.1086/160167}

\bibitem[{{Chevalier}(1998)}]{Chevalier1998}
---. 1998, \apj, 499, 810, \dodoi{10.1086/305676}

\bibitem[{{Chevalier} \& {Fransson}(1994)}]{Chevalier1994}
{Chevalier}, R.~A., \& {Fransson}, C. 1994, \apj, 420, 268,
  \dodoi{10.1086/173557}

\bibitem[{{Chevalier} \& {Fransson}(2006)}]{Chevalier2006}
---. 2006, \apj, 651, 381, \dodoi{10.1086/507606}

\bibitem[{{Chomiuk} {et~al.}(2012){Chomiuk}, {Soderberg}, {Moe}, {Chevalier},
  {Rupen}, {Badenes}, {Margutti}, {Fransson}, {Fong}, \&
  {Dittmann}}]{Chomiuk2012}
{Chomiuk}, L., {Soderberg}, A.~M., {Moe}, M., {et~al.} 2012, \apj, 750, 164,
  \dodoi{10.1088/0004-637X/750/2/164}

\bibitem[{{Chomiuk} {et~al.}(2016){Chomiuk}, {Soderberg}, {Chevalier},
  {Bruzewski}, {Foley}, {Parrent}, {Strader}, {Badenes}, {Fransson}, {Kamble},
  {Margutti}, {Rupen}, \& {Simon}}]{Chomiuk2015}
{Chomiuk}, L., {Soderberg}, A.~M., {Chevalier}, R.~A., {et~al.} 2016, \apj,
  821, 119, \dodoi{10.3847/0004-637X/821/2/119}

\bibitem[{{Chugai}(2008)}]{Chugai2008}
{Chugai}, N.~N. 2008, Astronomy Letters, 34, 389,
  \dodoi{10.1134/S1063773708060030}

\bibitem[{{Condon} {et~al.}(1998){Condon}, {Cotton}, {Greisen}, {Yin},
  {Perley}, {Taylor}, \& {Broderick}}]{Condon1998}
{Condon}, J.~J., {Cotton}, W.~D., {Greisen}, E.~W., {et~al.} 1998, \aj, 115,
  1693, \dodoi{10.1086/300337}

\bibitem[{{Cornwell}(2008)}]{Cornwell2008}
{Cornwell}, T.~J. 2008, IEEE Journal of Selected Topics in Signal Processing,
  2, 793, \dodoi{10.1109/JSTSP.2008.2006388}

\bibitem[{{Cowan} \& {Branch}(1982)}]{Cowan1982}
{Cowan}, J.~J., \& {Branch}, D. 1982, \apj, 258, 31, \dodoi{10.1086/160046}

\bibitem[{{Dan} {et~al.}(2011){Dan}, {Rosswog}, {Guillochon}, \&
  {Ramirez-Ruiz}}]{Dan2011}
{Dan}, M., {Rosswog}, S., {Guillochon}, J., \& {Ramirez-Ruiz}, E. 2011, \apj,
  737, 89, \dodoi{10.1088/0004-637X/737/2/89}

\bibitem[{{Danehkar} {et~al.}(2012){Danehkar}, {Frew}, {Parker}, \& {De
  Marco}}]{Danehkar2012}
{Danehkar}, A., {Frew}, D.~J., {Parker}, Q.~A., \& {De Marco}, O. 2012, in IAU
  Symposium, Vol. 283, IAU Symposium, 340--341

\bibitem[{{Darnley} {et~al.}(2017){Darnley}, {Hounsell}, {Godon}, {Perley},
  {Henze}, {Kuin}, {Williams}, {Williams}, {Bode}, {Harman}, {Hornoch}, {Link},
  {Ness}, {Ribeiro}, {Sion}, {Shafter}, \& {Shara}}]{Darnley2017}
{Darnley}, M.~J., {Hounsell}, R., {Godon}, P., {et~al.} 2017, \apj, 849, 96,
  \dodoi{10.3847/1538-4357/aa9062}

\bibitem[{{Darnley} {et~al.}(2019){Darnley}, {Hounsell}, {O'Brien}, {Henze},
  {Rodr{\'{\i}}guez-Gil}, {Shafter}, {Shara}, {Vaytet}, {Bode}, {Ciardullo},
  {Davis}, {Galera-Rosillo}, {Harman}, {Harvey}, {Healy}, {Ness}, {Ribeiro}, \&
  {Williams}}]{Darnley2019}
{Darnley}, M.~J., {Hounsell}, R., {O'Brien}, T.~J., {et~al.} 2019, \nat, 565,
  460, \dodoi{10.1038/s41586-018-0825-4}

\bibitem[{{de Vaucouleurs} \& {Corwin}(1985)}]{deVauc1985}
{de Vaucouleurs}, G., \& {Corwin}, H.~G., J. 1985, \apj, 295, 287,
  \dodoi{10.1086/163374}

\bibitem[{{DeLaney} {et~al.}(2002){DeLaney}, {Koralesky}, {Rudnick}, \&
  {Dickel}}]{DeLaney2002}
{DeLaney}, T., {Koralesky}, B., {Rudnick}, L., \& {Dickel}, J.~R. 2002, \apj,
  580, 914, \dodoi{10.1086/343787}

\bibitem[{{Deufel} {et~al.}(1999){Deufel}, {Barwig}, {{\v{S}}imi{\'c} },
  {Wolf}, \& {Drory}}]{Deufel1999}
{Deufel}, B., {Barwig}, H., {{\v{S}}imi{\'c} }, D., {Wolf}, S., \& {Drory}, N.
  1999, \aap, 343, 455.
\newblock \doarXiv{astro-ph/9811285}

\bibitem[{{Dimitriadis} {et~al.}(2014){Dimitriadis}, {Chiotellis}, \&
  {Vink}}]{Dimitriadis2014}
{Dimitriadis}, G., {Chiotellis}, A., \& {Vink}, J. 2014, \mnras, 443, 1370,
  \dodoi{10.1093/mnras/stu1249}

\bibitem[{{Dubner} \& {Giacani}(2015)}]{Dubner2015}
{Dubner}, G., \& {Giacani}, E. 2015, \aapr, 23, 3,
  \dodoi{10.1007/s00159-015-0083-5}

\bibitem[{{Eck} {et~al.}(2002){Eck}, {Cowan}, \& {Branch}}]{Eck2002}
{Eck}, C.~R., {Cowan}, J.~J., \& {Branch}, D. 2002, \apj, 573, 306,
  \dodoi{10.1086/340583}

\bibitem[{{Ferri{\`e}re}(2001)}]{Ferriere2001}
{Ferri{\`e}re}, K.~M. 2001, Reviews of Modern Physics, 73, 1031,
  \dodoi{10.1103/RevModPhys.73.1031}

\bibitem[{{Fesen} {et~al.}(2016){Fesen}, {Hoeflich}, \& {Hamilton}}]{Fesen2016}
{Fesen}, R., {Hoeflich}, P., \& {Hamilton}, A. 2016, in Supernova Remnants: An
  Odyssey in Space after Stellar Death, 115

\bibitem[{{Freedman} {et~al.}(2001){Freedman}, {Madore}, {Gibson}, {Ferrarese},
  {Kelson}, {Sakai}, {Mould}, {Kennicutt}, {Ford}, {Graham}, {Huchra},
  {Hughes}, {Illingworth}, {Macri}, \& {Stetson}}]{Freedman2001}
{Freedman}, W.~L., {Madore}, B.~F., {Gibson}, B.~K., {et~al.} 2001, \apj, 553,
  47, \dodoi{10.1086/320638}

\bibitem[{{Glasner} {et~al.}(2018){Glasner}, {Livne}, {Steinberg},
  {Yalinewich}, \& {Truran}}]{Glasner2018}
{Glasner}, A., {Livne}, E., {Steinberg}, E., {Yalinewich}, A., \& {Truran},
  J.~W. 2018, arXiv e-prints, arXiv:1803.00941.
\newblock \doarXiv{1803.00941}

\bibitem[{{Graham} {et~al.}(2019){Graham}, {Harris}, {Nugent}, {Maguire},
  {Sullivan}, {Smith}, {Valenti}, {Goobar}, {Fox}, {Shen}, {Kelly}, {McCully},
  {Brink}, \& {Filippenko}}]{Graham2019}
{Graham}, M.~L., {Harris}, C.~E., {Nugent}, P.~E., {et~al.} 2019, \apj, 871,
  62, \dodoi{10.3847/1538-4357/aaf41e}

\bibitem[{{Green}(2014)}]{Green2014}
{Green}, D.~A. 2014, Bulletin of the Astronomical Society of India, 42, 47.
\newblock \doarXiv{1409.0637}

\bibitem[{{Green} {et~al.}(2008){Green}, {Reynolds}, {Borkowski}, {Hwang},
  {Harrus}, \& {Petre}}]{Green2008}
{Green}, D.~A., {Reynolds}, S.~P., {Borkowski}, K.~J., {et~al.} 2008, \mnras,
  387, L54, \dodoi{10.1111/j.1745-3933.2008.00484.x}

\bibitem[{{Guillochon} {et~al.}(2010){Guillochon}, {Dan}, {Ramirez-Ruiz}, \&
  {Rosswog}}]{Guillochon2010}
{Guillochon}, J., {Dan}, M., {Ramirez-Ruiz}, E., \& {Rosswog}, S. 2010, \apjl,
  709, L64, \dodoi{10.1088/2041-8205/709/1/L64}

\bibitem[{{Hachisu} {et~al.}(1996){Hachisu}, {Kato}, \& {Nomoto}}]{Hachisu1996}
{Hachisu}, I., {Kato}, M., \& {Nomoto}, K. 1996, \apjl, 470, L97,
  \dodoi{10.1086/310303}

\bibitem[{{Hancock} {et~al.}(2011){Hancock}, {Gaensler}, \&
  {Murphy}}]{Hancock2011}
{Hancock}, P.~J., {Gaensler}, B.~M., \& {Murphy}, T. 2011, \apjl, 735, L35,
  \dodoi{10.1088/2041-8205/735/2/L35}

\bibitem[{Harris {et~al.}(2016)Harris, Nugent, \& Kasen}]{Harris2016}
Harris, C.~E., Nugent, P.~E., \& Kasen, D.~N. 2016, arXiv.org, 100

\bibitem[{{Hillebrandt} \& {Niemeyer}(2000)}]{Hillebrandt2000}
{Hillebrandt}, W., \& {Niemeyer}, J.~C. 2000, \araa, 38, 191,
  \dodoi{10.1146/annurev.astro.38.1.191}

\bibitem[{{Holmbo} {et~al.}(2018){Holmbo}, {Stritzinger}, {Shappee}, {Tucker},
  {Zheng}, {Ashall}, {Phillips}, {Contreras}, {Filippenko}, {Hoeflich},
  {Huber}, {Wang}, {Zhang}, {Anais}, {Baron}, {Burns}, {Campillay},
  {Castellon}, {Corco}, {Hsiao}, {Krisciunas}, {Morrell}, {Nielsen}, {Persson},
  {Piro}, {Taddia}, {Tomasella}, {Zhang}, \& {Zhao}}]{Holmbo2018}
{Holmbo}, S., {Stritzinger}, M.~D., {Shappee}, B.~J., {et~al.} 2018, arXiv
  e-prints.
\newblock \doarXiv{1809.01359}

\bibitem[{{Horesh} {et~al.}(2012){Horesh}, {Kulkarni}, {Fox}, {Carpenter},
  {Kasliwal}, {Ofek}, {Quimby}, {Gal-Yam}, {Cenko}, {de Bruyn}, {Kamble},
  {Wijers}, {van der Horst}, {Kouveliotou}, {Podsiadlowski}, {Sullivan},
  {Maguire}, {Howell}, {Nugent}, {Gehrels}, {Law}, {Poznanski}, \&
  {Shara}}]{Horesh2012}
{Horesh}, A., {Kulkarni}, S.~R., {Fox}, D.~B., {et~al.} 2012, \apj, 746, 21,
  \dodoi{10.1088/0004-637X/746/1/21}

\bibitem[{{Horesh} {et~al.}(2013){Horesh}, {Stockdale}, {Fox}, {Frail},
  {Carpenter}, {Kulkarni}, {Ofek}, {Gal-Yam}, {Kasliwal}, {Arcavi}, {Quimby},
  {Cenko}, {Nugent}, {Bloom}, {Law}, {Poznanski}, {Gorbikov}, {Polishook},
  {Yaron}, {Ryder}, {Weiler}, {Bauer}, {Van Dyk}, {Immler}, {Panagia},
  {Pooley}, \& {Kassim}}]{Horesh2013}
{Horesh}, A., {Stockdale}, C., {Fox}, D.~B., {et~al.} 2013, \mnras, 436, 1258,
  \dodoi{10.1093/mnras/stt1645}

\bibitem[{{Hunter}(2007)}]{Hunter2007}
{Hunter}, J.~D. 2007, Computing in Science and Engineering, 9, 90,
  \dodoi{10.1109/MCSE.2007.55}

\bibitem[{{Iben} \& {Tutukov}(1984)}]{Iben1984}
{Iben}, I., J., \& {Tutukov}, A.~V. 1984, \apjs, 54, 335,
  \dodoi{10.1086/190932}

\bibitem[{{Ilkov} \& {Soker}(2012)}]{Ilkov2012}
{Ilkov}, M., \& {Soker}, N. 2012, \mnras, 419, 1695,
  \dodoi{10.1111/j.1365-2966.2011.19833.x}

\bibitem[{{Ivanova} {et~al.}(2013){Ivanova}, {Justham}, {Chen}, {De Marco},
  {Fryer}, {Gaburov}, {Ge}, {Glebbeek}, {Han}, {Li}, {Lu}, {Marsh},
  {Podsiadlowski}, {Potter}, {Soker}, {Taam}, {Tauris}, {van den Heuvel}, \&
  {Webbink}}]{Ivanova2013}
{Ivanova}, N., {Justham}, S., {Chen}, X., {et~al.} 2013, Astronomy and
  Astrophysics Review, 21, 59, \dodoi{10.1007/s00159-013-0059-2}

\bibitem[{{Jacoby} {et~al.}(2001){Jacoby}, {Ferland}, \&
  {Korista}}]{Jacoby2001}
{Jacoby}, G.~H., {Ferland}, G.~J., \& {Korista}, K.~T. 2001, \apj, 560, 272,
  \dodoi{10.1086/322489}

\bibitem[{{Jarrett}(1973)}]{Jarrett1973}
{Jarrett}, A.~H. 1973, Information Bulletin on Variable Stars, 828

\bibitem[{{Ji} {et~al.}(2013){Ji}, {Fisher}, {Garc{\'\i}a-Berro}, {Tzeferacos},
  {Jordan}, {Lee}, {Lor{\'e}n-Aguilar}, {Cremer}, \& {Behrends}}]{Ji2013}
{Ji}, S., {Fisher}, R.~T., {Garc{\'\i}a-Berro}, E., {et~al.} 2013, \apj, 773,
  136, \dodoi{10.1088/0004-637X/773/2/136}

\bibitem[{{Kasen}(2010)}]{Kasen2010}
{Kasen}, D. 2010, \apj, 708, 1025, \dodoi{10.1088/0004-637X/708/2/1025}

\bibitem[{{Kashi} \& {Soker}(2011)}]{Kashi2011}
{Kashi}, A., \& {Soker}, N. 2011, \mnras, 417, 1466,
  \dodoi{10.1111/j.1365-2966.2011.19361.x}

\bibitem[{{Kirshner} \& {Oke}(1975)}]{Kirshner1975}
{Kirshner}, R.~P., \& {Oke}, J.~B. 1975, \apj, 200, 574, \dodoi{10.1086/153824}

\bibitem[{{Kobulnicky} \& {Skillman}(1995)}]{Kobulnicky1995}
{Kobulnicky}, H.~A., \& {Skillman}, E.~D. 1995, \apjl, 454, L121,
  \dodoi{10.1086/309791}

\bibitem[{{Kollmeier} {et~al.}(2019){Kollmeier}, {Chen}, {Dong}, {Morrell},
  {Phillips}, {Kushnir}, {Prieto}, {Piro}, \& {Simon}}]{Kollmeier2019}
{Kollmeier}, J.~A., {Chen}, P., {Dong}, S., {et~al.} 2019, \mnras, 486, 3041,
  \dodoi{10.1093/mnras/stz953}

\bibitem[{{Kozlova} \& {Blinnikov}(2018)}]{Kozlova2018}
{Kozlova}, A.~V., \& {Blinnikov}, S.~I. 2018, in Journal of Physics Conference
  Series, Vol. 1038, Journal of Physics Conference Series, 012006

\bibitem[{{Kromer} {et~al.}(2010){Kromer}, {Sim}, {Fink}, {R{\"o}pke},
  {Seitenzahl}, \& {Hillebrandt}}]{Kromer2010}
{Kromer}, M., {Sim}, S.~A., {Fink}, M., {et~al.} 2010, \apj, 719, 1067,
  \dodoi{10.1088/0004-637X/719/2/1067}

\bibitem[{{Kushnir} {et~al.}(2013){Kushnir}, {Katz}, {Dong}, {Livne}, \&
  {Fern{\'a}ndez}}]{Kushnir2013}
{Kushnir}, D., {Katz}, B., {Dong}, S., {Livne}, E., \& {Fern{\'a}ndez}, R.
  2013, \apj, 778, L37, \dodoi{10.1088/2041-8205/778/2/L37}

\bibitem[{{Leibundgut} {et~al.}(1991){Leibundgut}, {Tammann}, {Cadonau}, \&
  {Cerrito}}]{Leibundgut1991}
{Leibundgut}, B., {Tammann}, G.~A., {Cadonau}, R., \& {Cerrito}, D. 1991,
  \aaps, 89, 537

\bibitem[{{Lindblom}(1999)}]{Lindblom1999}
{Lindblom}, L. 1999, \prd, 60, 064007, \dodoi{10.1103/PhysRevD.60.064007}

\bibitem[{{Liu} {et~al.}(2018){Liu}, {Wang}, \& {Han}}]{Liu2018}
{Liu}, D., {Wang}, B., \& {Han}, Z. 2018, \mnras, 473, 5352,
  \dodoi{10.1093/mnras/stx2756}

\bibitem[{{Livio} \& {Riess}(2003)}]{Livio2003}
{Livio}, M., \& {Riess}, A.~G. 2003, \apjl, 594, L93, \dodoi{10.1086/378765}

\bibitem[{{Livne} \& {Arnett}(1995)}]{Livne1995}
{Livne}, E., \& {Arnett}, D. 1995, \apj, 452, 62, \dodoi{10.1086/176279}

\bibitem[{{MacLeod} {et~al.}(2017){MacLeod}, {Macias}, {Ramirez-Ruiz},
  {Grindlay}, {Batta}, \& {Montes}}]{MacLeod2017}
{MacLeod}, M., {Macias}, P., {Ramirez-Ruiz}, E., {et~al.} 2017, \apj, 835, 282,
  \dodoi{10.3847/1538-4357/835/2/282}

\bibitem[{{Maguire} {et~al.}(2013){Maguire}, {Sullivan}, {Patat}, {Gal-Yam},
  {Hook}, {Dhawan}, {Howell}, {Mazzali}, {Nugent}, {Pan}, {Podsiadlowski},
  {Simon}, {Sternberg}, {Valenti}, {Baltay}, {Bersier}, {Blagorodnova}, {Chen},
  {Ellman}, {Feindt}, {F{\"o}rster}, {Fraser}, {Gonz{\'a}lez-Gait{\'a}n},
  {Graham}, {Guti{\'e}rrez}, {Hachinger}, {Hadjiyska}, {Inserra}, {Knapic},
  {Laher}, {Leloudas}, {Margheim}, {McKinnon}, {Molinaro}, {Morrell}, {Ofek},
  {Rabinowitz}, {Rest}, {Sand}, {Smareglia}, {Smartt}, {Taddia}, {Walker},
  {Walton}, \& {Young}}]{Maguire2013}
{Maguire}, K., {Sullivan}, M., {Patat}, F., {et~al.} 2013, \mnras, 436, 222,
  \dodoi{10.1093/mnras/stt1586}

\bibitem[{{Maoz} {et~al.}(2014){Maoz}, {Mannucci}, \& {Nelemans}}]{Maoz2014}
{Maoz}, D., {Mannucci}, F., \& {Nelemans}, G. 2014, \araa, 52, 107,
  \dodoi{10.1146/annurev-astro-082812-141031}

\bibitem[{{Margutti} {et~al.}(2014){Margutti}, {Parrent}, {Kamble},
  {Soderberg}, {Foley}, {Milisavljevic}, {Drout}, \& {Kirshner}}]{Margutti2014}
{Margutti}, R., {Parrent}, J., {Kamble}, A., {et~al.} 2014, \apj, 790, 52,
  \dodoi{10.1088/0004-637X/790/1/52}

\bibitem[{{Margutti} {et~al.}(2012){Margutti}, {Soderberg}, {Chomiuk},
  {Chevalier}, {Hurley}, {Milisavljevic}, {Foley}, {Hughes}, {Slane},
  {Fransson}, {Moe}, {Barthelmy}, {Boynton}, {Briggs}, {Connaughton}, {Costa},
  {Cummings}, {Del Monte}, {Enos}, {Fellows}, {Feroci}, {Fukazawa}, {Gehrels},
  {Goldsten}, {Golovin}, {Hanabata}, {Harshman}, {Krimm}, {Litvak},
  {Makishima}, {Marisaldi}, {Mitrofanov}, {Murakami}, {Ohno}, {Palmer},
  {Sanin}, {Starr}, {Svinkin}, {Takahashi}, {Tashiro}, {Terada}, \&
  {Yamaoka}}]{Margutti2012}
{Margutti}, R., {Soderberg}, A.~M., {Chomiuk}, L., {et~al.} 2012, \apj, 751,
  134, \dodoi{10.1088/0004-637X/751/2/134}

\bibitem[{Mattila {et~al.}(2010)Mattila, Lundqvist, Gr{\"o}ningsson, Meikle,
  Stathakis, Fransson, \& Cannon}]{Mattila2010}
Mattila, S., Lundqvist, P., Gr{\"o}ningsson, P., {et~al.} 2010, The
  Astrophysical Journal, 717, 1140

\bibitem[{{Matzner} \& {McKee}(1999)}]{Matzner1999}
{Matzner}, C.~D., \& {McKee}, C.~F. 1999, \apj, 510, 379,
  \dodoi{10.1086/306571}

\bibitem[{{McMullin} {et~al.}(2007){McMullin}, {Waters}, {Schiebel}, {Young},
  \& {Golap}}]{McMullin2007}
{McMullin}, J.~P., {Waters}, B., {Schiebel}, D., {Young}, W., \& {Golap}, K.
  2007, in Astronomical Society of the Pacific Conference Series, Vol. 376,
  Astronomical Data Analysis Software and Systems XVI, ed. R.~A. {Shaw},
  F.~{Hill}, \& D.~J. {Bell}, 127

\bibitem[{{Monreal-Ibero} {et~al.}(2010){Monreal-Ibero}, {V{\'{\i}}lchez},
  {Walsh}, \& {Mu{\~n}oz-Tu{\~n}{\'o}n}}]{Monreal2010}
{Monreal-Ibero}, A., {V{\'{\i}}lchez}, J.~M., {Walsh}, J.~R., \&
  {Mu{\~n}oz-Tu{\~n}{\'o}n}, C. 2010, \aap, 517, A27,
  \dodoi{10.1051/0004-6361/201014154}

\bibitem[{{Moore} \& {Bildsten}(2012)}]{Moore2012}
{Moore}, K., \& {Bildsten}, L. 2012, \apj, 761, 182,
  \dodoi{10.1088/0004-637X/761/2/182}

\bibitem[{{Morlino} \& {Caprioli}(2012)}]{Morlino2012}
{Morlino}, G., \& {Caprioli}, D. 2012, \aap, 538, A81,
  \dodoi{10.1051/0004-6361/201117855}

\bibitem[{{Munari} {et~al.}(1999){Munari}, {Zwitter}, {Tomov}, {Bonifacio},
  {Molaro}, {Selvelli}, {Tomasella}, {Niedzielski}, \& {Pearce}}]{Munari1999}
{Munari}, U., {Zwitter}, T., {Tomov}, T., {et~al.} 1999, \aap, 347, L39

\bibitem[{{Nomoto}(1982)}]{Nomoto1982}
{Nomoto}, K. 1982, \apj, 253, 798, \dodoi{10.1086/159682}

\bibitem[{{Nomoto} {et~al.}(1984){Nomoto}, {Thielemann}, \&
  {Yokoi}}]{Nomoto1984}
{Nomoto}, K., {Thielemann}, F.~K., \& {Yokoi}, K. 1984, \apj, 286, 644,
  \dodoi{10.1086/162639}

\bibitem[{Offringa {et~al.}(2012)Offringa, van~de Gronde, \&
  Roerdink}]{Offringa2012}
Offringa, A.~R., van~de Gronde, J.~J., \& Roerdink, J. B. T.~M. 2012, A\&A, 539

\bibitem[{{Panagia} {et~al.}(2006){Panagia}, {Van Dyk}, {Weiler}, {Sramek},
  {Stockdale}, \& {Murata}}]{Panagia2006}
{Panagia}, N., {Van Dyk}, S.~D., {Weiler}, K.~W., {et~al.} 2006, \apj, 646,
  369, \dodoi{10.1086/504710}

\bibitem[{{Patat} {et~al.}(2011){Patat}, {Chugai}, {Podsiadlowski}, {Mason},
  {Melo}, \& {Pasquini}}]{Patat2011}
{Patat}, F., {Chugai}, N.~N., {Podsiadlowski}, P., {et~al.} 2011, \aap, 530,
  A63, \dodoi{10.1051/0004-6361/201116865}

\bibitem[{{Patat} {et~al.}(2007){Patat}, {Chandra}, {Chevalier}, {Justham},
  {Podsiadlowski}, {Wolf}, {Gal-Yam}, {Pasquini}, {Crawford}, {Mazzali},
  {Pauldrach}, {Nomoto}, {Benetti}, {Cappellaro}, {Elias-Rosa}, {Hillebrandt},
  {Leonard}, {Pastorello}, {Renzini}, {Sabbadin}, {Simon}, \&
  {Turatto}}]{Patat2007}
{Patat}, F., {Chandra}, P., {Chevalier}, R., {et~al.} 2007, Science, 317, 924,
  \dodoi{10.1126/science.1143005}

\bibitem[{{Perley} \& {Butler}(2017)}]{Perley2017}
{Perley}, R.~A., \& {Butler}, B.~J. 2017, \apjs, 230, 7,
  \dodoi{10.3847/1538-4365/aa6df9}

\bibitem[{{Perlmutter} {et~al.}(1999){Perlmutter}, {Aldering}, {Goldhaber},
  {Knop}, {Nugent}, {Castro}, {Deustua}, {Fabbro}, {Goobar}, {Groom}, {Hook},
  {Kim}, {Kim}, {Lee}, {Nunes}, {Pain}, {Pennypacker}, {Quimby}, {Lidman},
  {Ellis}, {Irwin}, {McMahon}, {Ruiz-Lapuente}, {Walton}, {Schaefer}, {Boyle},
  {Filippenko}, {Matheson}, {Fruchter}, {Panagia}, {Newberg}, {Couch}, \&
  {Project}}]{Perlmutter1999}
{Perlmutter}, S., {Aldering}, G., {Goldhaber}, G., {et~al.} 1999, \apj, 517,
  565, \dodoi{10.1086/307221}

\bibitem[{{Pickering}(1895)}]{Pickering1895}
{Pickering}, E.~C. 1895, Harvard College Observatory Circular, 4, 1

\bibitem[{{Raskin} \& {Kasen}(2013)}]{Raskin2013}
{Raskin}, C., \& {Kasen}, D. 2013, \apj, 772, 1,
  \dodoi{10.1088/0004-637X/772/1/1}

\bibitem[{{Rau} \& {Cornwell}(2011)}]{Rau2011}
{Rau}, U., \& {Cornwell}, T.~J. 2011, \aap, 532, A71,
  \dodoi{10.1051/0004-6361/201117104}

\bibitem[{{Reynolds} {et~al.}(2008){Reynolds}, {Borkowski}, {Green}, {Hwang},
  {Harrus}, \& {Petre}}]{Reynolds2008}
{Reynolds}, S.~P., {Borkowski}, K.~J., {Green}, D.~A., {et~al.} 2008, \apjl,
  680, L41, \dodoi{10.1086/589570}

\bibitem[{{Reynolds} {et~al.}(2007){Reynolds}, {Borkowski}, {Hwang}, {Hughes},
  {Badenes}, {Laming}, \& {Blondin}}]{Reynolds2007}
{Reynolds}, S.~P., {Borkowski}, K.~J., {Hwang}, U., {et~al.} 2007, \apjl, 668,
  L135, \dodoi{10.1086/522830}

\bibitem[{Reynoso \& Goss(1999)}]{Reynoso1999}
Reynoso, E.~M., \& Goss, W.~M. 1999, The Astronomical Journal, 118, 926

\bibitem[{{Riess} {et~al.}(1998){Riess}, {Filippenko}, {Challis},
  {Clocchiatti}, {Diercks}, {Garnavich}, {Gilliland}, {Hogan}, {Jha},
  {Kirshner}, {Leibundgut}, {Phillips}, {Reiss}, {Schmidt}, {Schommer},
  {Smith}, {Spyromilio}, {Stubbs}, {Suntzeff}, \& {Tonry}}]{Riess1998}
{Riess}, A.~G., {Filippenko}, A.~V., {Challis}, P., {et~al.} 1998, \aj, 116,
  1009, \dodoi{10.1086/300499}

\bibitem[{{Roth} \& {Kasen}(2015)}]{Roth2015}
{Roth}, N., \& {Kasen}, D. 2015, \apjs, 217, 9,
  \dodoi{10.1088/0067-0049/217/1/9}

\bibitem[{{Ruiter} {et~al.}(2011){Ruiter}, {Belczynski}, {Sim}, {Hillebrand t},
  {Fryer}, {Fink}, \& {Kromer}}]{Ruiter2011}
{Ruiter}, A.~J., {Belczynski}, K., {Sim}, S.~A., {et~al.} 2011, \mnras, 417,
  408, \dodoi{10.1111/j.1365-2966.2011.19276.x}

\bibitem[{{Ruiter} {et~al.}(2013){Ruiter}, {Sim}, {Pakmor}, {Kromer},
  {Seitenzahl}, {Belczynski}, {Fink}, {Herzog}, {Hillebrandt}, {R{\"o}pke}, \&
  {Taubenberger}}]{Ruiter2013}
{Ruiter}, A.~J., {Sim}, S.~A., {Pakmor}, R., {et~al.} 2013, \mnras, 429, 1425,
  \dodoi{10.1093/mnras/sts423}

\bibitem[{{Ruiz-Lapuente}(2004)}]{RuizLapuente2004}
{Ruiz-Lapuente}, P. 2004, \apj, 612, 357, \dodoi{10.1086/422419}

\bibitem[{{Russell} \& {Immler}(2012)}]{Russell2012}
{Russell}, B.~R., \& {Immler}, S. 2012, \apjl, 748, L29,
  \dodoi{10.1088/2041-8205/748/2/L29}

\bibitem[{{Sarbadhicary} {et~al.}(2017){Sarbadhicary}, {Badenes}, {Chomiuk},
  {Caprioli}, \& {Huizenga}}]{Sarbad2016}
{Sarbadhicary}, S.~K., {Badenes}, C., {Chomiuk}, L., {Caprioli}, D., \&
  {Huizenga}, D. 2017, \mnras, 464, 2326, \dodoi{10.1093/mnras/stw2566}

\bibitem[{{Sarbadhicary} {et~al.}(2019{\natexlab{a}}){Sarbadhicary}, {Badenes},
  {Chomiuk}, {Caprioli}, \& {Huizenga}}]{Sarbadhicary2019}
---. 2019{\natexlab{a}}, \mnras, 1400, \dodoi{10.1093/mnras/stz1490}

\bibitem[{{Sarbadhicary} {et~al.}(2019{\natexlab{b}}){Sarbadhicary}, {Chomiuk},
  {Badenes}, {Tremou}, {Soderberg}, \& {Sjouwerman}}]{Sarbad2017}
{Sarbadhicary}, S.~K., {Chomiuk}, L., {Badenes}, C., {et~al.}
  2019{\natexlab{b}}, \apj, 872, 191, \dodoi{10.3847/1538-4357/ab027f}

\bibitem[{{Scalzo} {et~al.}(2019){Scalzo}, {Parent}, {Burns}, {Childress},
  {Tucker}, {Brown}, {Contreras}, {Hsiao}, {Krisciunas}, {Morrell}, {Phillips},
  {Piro}, {Stritzinger}, \& {Suntzeff}}]{Scalzo2019}
{Scalzo}, R.~A., {Parent}, E., {Burns}, C., {et~al.} 2019, \mnras, 483, 628,
  \dodoi{10.1093/mnras/sty3178}

\bibitem[{{Schaefer}(1995)}]{Schaefer1995}
{Schaefer}, B.~E. 1995, \apjl, 447, L13, \dodoi{10.1086/309549}

\bibitem[{{Schwab} {et~al.}(2016){Schwab}, {Quataert}, \& {Kasen}}]{Schwab2016}
{Schwab}, J., {Quataert}, E., \& {Kasen}, D. 2016, \mnras, 463, 3461,
  \dodoi{10.1093/mnras/stw2249}

\bibitem[{{Seaquist} \& {Taylor}(1990)}]{Seaquist1990}
{Seaquist}, E.~R., \& {Taylor}, A.~R. 1990, \apj, 349, 313,
  \dodoi{10.1086/168315}

\bibitem[{{Shen} {et~al.}(2012){Shen}, {Bildsten}, {Kasen}, \&
  {Quataert}}]{Shen2012}
{Shen}, K.~J., {Bildsten}, L., {Kasen}, D., \& {Quataert}, E. 2012, \apj, 748,
  35, \dodoi{10.1088/0004-637X/748/1/35}

\bibitem[{{Shen} {et~al.}(2013){Shen}, {Guillochon}, \& {Foley}}]{Shen2013}
{Shen}, K.~J., {Guillochon}, J., \& {Foley}, R.~J. 2013, \apj, 770, L35,
  \dodoi{10.1088/2041-8205/770/2/L35}

\bibitem[{{Shen} {et~al.}(2018){Shen}, {Kasen}, {Miles}, \&
  {Townsley}}]{Shen2018}
{Shen}, K.~J., {Kasen}, D., {Miles}, B.~J., \& {Townsley}, D.~M. 2018, \apj,
  854, 52, \dodoi{10.3847/1538-4357/aaa8de}

\bibitem[{{Silverman} {et~al.}(2013){Silverman}, {Nugent}, {Gal-Yam},
  {Sullivan}, {Howell}, {Filippenko}, {Arcavi}, {Ben-Ami}, {Bloom}, {Cenko},
  {Cao}, {Chornock}, {Clubb}, {Coil}, {Foley}, {Graham}, {Griffith}, {Horesh},
  {Kasliwal}, {Kulkarni}, {Leonard}, {Li}, {Matheson}, {Miller}, {Modjaz},
  {Ofek}, {Pan}, {Perley}, {Poznanski}, {Quimby}, {Steele}, {Sternberg}, {Xu},
  \& {Yaron}}]{Silverman2013}
{Silverman}, J.~M., {Nugent}, P.~E., {Gal-Yam}, A., {et~al.} 2013, \apjs, 207,
  3, \dodoi{10.1088/0067-0049/207/1/3}

\bibitem[{{Sim} {et~al.}(2012){Sim}, {Fink}, {Kromer}, {R{\"o}pke}, {Ruiter},
  \& {Hillebrandt}}]{Sim2012}
{Sim}, S.~A., {Fink}, M., {Kromer}, M., {et~al.} 2012, \mnras, 420, 3003,
  \dodoi{10.1111/j.1365-2966.2011.20162.x}

\bibitem[{{Simon} {et~al.}(2009){Simon}, {Gal-Yam}, {Gnat}, {Quimby},
  {Ganeshalingam}, {Silverman}, {Blondin}, {Li}, {Filippenko}, {Wheeler},
  {Kirshner}, {Patat}, {Nugent}, {Foley}, {Vogt}, {Butler}, {Peek},
  {Rosolowsky}, {Herczeg}, {Sauer}, \& {Mazzali}}]{Simon2009}
{Simon}, J.~D., {Gal-Yam}, A., {Gnat}, O., {et~al.} 2009, \apj, 702, 1157,
  \dodoi{10.1088/0004-637X/702/2/1157}

\bibitem[{{Soker}(2011)}]{Soker2011}
{Soker}, N. 2011, arXiv e-prints, arXiv:1109.4652.
\newblock \doarXiv{1109.4652}

\bibitem[{{Sternberg} {et~al.}(2011){Sternberg}, {Gal-Yam}, {Simon}, {Leonard},
  {Quimby}, {Phillips}, {Morrell}, {Thompson}, {Ivans}, {Marshall},
  {Filippenko}, {Marcy}, {Bloom}, {Patat}, {Foley}, {Yong}, {Penprase},
  {Beeler}, {Allende Prieto}, \& {Stringfellow}}]{Sternberg2011}
{Sternberg}, A., {Gal-Yam}, A., {Simon}, J.~D., {et~al.} 2011, Science, 333,
  856, \dodoi{10.1126/science.1203836}

\bibitem[{Summers {et~al.}(2004)Summers, Stevens, Strickland, \&
  Heckman}]{Summers2004}
Summers, L.~K., Stevens, I.~R., Strickland, D.~K., \& Heckman, T.~M. 2004,
  Monthly Notices of the Royal Astronomical Society, 351, 1,
  \dodoi{10.1111/j.1365-2966.2004.07749.x}

\bibitem[{{Thielemann} {et~al.}(1986){Thielemann}, {Nomoto}, \&
  {Yokoi}}]{Thielemann1986}
{Thielemann}, F.~K., {Nomoto}, K., \& {Yokoi}, K. 1986, \aap, 158, 17

\bibitem[{{Toal{\'a}} \& {Arthur}(2016)}]{Toala2016}
{Toal{\'a}}, J.~A., \& {Arthur}, S.~J. 2016, \mnras, 463, 4438,
  \dodoi{10.1093/mnras/stw2307}

\bibitem[{{Tout} {et~al.}(2008){Tout}, {Wickramasinghe}, {Liebert}, {Ferrario},
  \& {Pringle}}]{Tout2008}
{Tout}, C.~A., {Wickramasinghe}, D.~T., {Liebert}, J., {Ferrario}, L., \&
  {Pringle}, J.~E. 2008, \mnras, 387, 897,
  \dodoi{10.1111/j.1365-2966.2008.13291.x}

\bibitem[{{Trimble}(1982)}]{Trimble1982}
{Trimble}, V. 1982, Reviews of Modern Physics, 54, 1183,
  \dodoi{10.1103/RevModPhys.54.1183}

\bibitem[{{Truelove} \& {McKee}(1999)}]{Truelove1999}
{Truelove}, J.~K., \& {McKee}, C.~F. 1999, \apjs, 120, 299,
  \dodoi{10.1086/313176}

\bibitem[{{Tsebrenko} \& {Soker}(2015)}]{Tsebrenko2015}
{Tsebrenko}, D., \& {Soker}, N. 2015, \mnras, 447, 2568,
  \dodoi{10.1093/mnras/stu2567}

\bibitem[{{van den Bergh}(1980)}]{vandenbergh1980}
{van den Bergh}, S. 1980, \pasp, 92, 122, \dodoi{10.1086/130631}

\bibitem[{{Wang}(2018)}]{Wang2018}
{Wang}, B. 2018, Research in Astronomy and Astrophysics, 18, 049,
  \dodoi{10.1088/1674-4527/18/5/49}

\bibitem[{{Webbink}(1984)}]{Webbink1984}
{Webbink}, R.~F. 1984, \apj, 277, 355, \dodoi{10.1086/161701}

\bibitem[{{Williams} {et~al.}(2017){Williams}, {Gizis}, \&
  {Berger}}]{Williams2017}
{Williams}, P.~K.~G., {Gizis}, J.~E., \& {Berger}, E. 2017, \apj, 834, 117,
  \dodoi{10.3847/1538-4357/834/2/117}

\bibitem[{{Woosley} \& {Kasen}(2011)}]{Woosley2011}
{Woosley}, S.~E., \& {Kasen}, D. 2011, \apj, 734, 38,
  \dodoi{10.1088/0004-637X/734/1/38}

\bibitem[{{Woosley} \& {Weaver}(1994)}]{Woosley1994}
{Woosley}, S.~E., \& {Weaver}, T.~A. 1994, \apj, 423, 371,
  \dodoi{10.1086/173813}

\bibitem[{{Yoon} \& {Langer}(2005)}]{Yoon2005}
{Yoon}, S.~C., \& {Langer}, N. 2005, in American Institute of Physics
  Conference Series, Vol. 797, Interacting Binaries: Accretion, Evolution, and
  Outcomes, ed. L.~{Burderi}, L.~A. {Antonelli}, F.~{D'Antona}, T.~{di Salvo},
  G.~L. {Israel}, L.~{Piersanti}, A.~{Tornamb{\`e}}, \& O.~{Straniero},
  651--654

\end{thebibliography}

\end{document}